\documentclass[preprint, 11pt,preprintnumbers,amsmath,amssymb,nofootinbib]{revtex4}
\usepackage{amsmath, amssymb,graphicx}
\usepackage{subfig}
\usepackage[countmax]{subfloat}
\usepackage{epstopdf}
\usepackage {amssymb}
\newcommand{\nc}{\newcommand}
\nc{\ba}{\begin{eqnarray}} \nc{\ea}{\end{eqnarray}}
\newcommand\be{\begin{equation}}
\newcommand\ee{\end{equation}}
\nc{\D}{\overline{\mbox{D3}}}

\nc{\ga}{\gamma} \nc{\tnu}{\tilde{\nu}} \nc{\tmu}{\tilde{\mu}}

\nc{\x}{{\bf{x}}}

\input{epsf.sty}
\begin{document}

\date{\today}
\title{Observable Signatures of Inflaton Decays}
\author{Diana Battefeld$^{1)}$} 
\email{dbattefe(AT)astro.physik.uni-goettingen.de}
\author{Thorsten Battefeld$^{1)}$}
\email{tbattefe(AT)astro.physik.uni-goettingen.de}
\author{John T.~Giblin, Jr.$^{2,3)}$}
\email{giblinj(AT)kenyon.edu}
\author{Evan K. Pease$^{2)}$}
\email{peasee(AT)kenyon.edu}
\affiliation{1)  Institute for Astrophysics,
University of Goettingen,
Friedrich Hund Platz 1,
D-37077 Gottingen, Germany}
\affiliation{2) Department of Physics, Kenyon College, Gambier, OH 43022}
\affiliation{3)The Perimeter Institute for Theoretical Physics, 31 Caroline St N, Waterloo, ON  N2L 2Y5, CANADA}

\begin{abstract}
We numerically compute features in the power-spectrum that originate from the decay of fields during inflation. Using a simple, phenomenological, multi-field setup, we increase the number of fields from a few to thousands. Whenever a field decays, its associated potential energy is transferred into radiation, causing a jump in the equation of state parameter and mode mixing at the perturbed level. We observe discrete steps in the power-spectrum if the number of fields is low, in agreement with analytic arguments in the literature. These features become increasingly smeared out once many fields decay within a given Hubble time. In this regime we confirm the validity of the analytic approach to staggered inflation, which is based on a coarse-graining procedure. Our numerical approach bridges the aforementioned analytic treatments, and can be used in more complicated scenarios.  
\end{abstract}
\maketitle
\newpage


\section{Introduction}
The construction of inflationary models in string theory is an active research field, see \cite{HenryTye:2006uv,Cline:2006hu,Burgess:2007pz,McAllister:2007bg,Baumann:2009ni, Mazumdar:2010sa} for reviews. Early proposals often involved few dynamical degrees of freedom, i.e.~a single field driving inflation, while all others are meticulously stabilized as in the KKLMMT \cite{Kachru:2003sx} brane inflation \cite{dvali-tye,Alexander:2001ks,collection,Dvali:2001fw,Firouzjahi:2003zy,Burgess:2004kv,Buchel,Iizuka:2004ct} setup. These models offer computational control and predictability, but they appear simplistic in the absence of any {\sl a priory} reason or need for such frugality; for instance, if inflation is driven by a brane/anti-brane pair, why shouldn't more pairs be included? As a consequence, multi-field models have become increasingly popular \cite{Dimopoulos:2005ac,Cline:2005ty, Ashoorioon:2009wa,Ashoorioon:2009sr,Becker:2005sg,Ashoorioon:2006wc,Ashoorioon:2008qr,Piao:2002vf,Majumdar:2003kd,Battefeld:2010rf,Firouzjahi:2010ga}, especially since observable non-Gaussianities, for which there is emerging experimental evidence \cite{Komatsu:2010fb}, are possible \cite{Senatore:2010wk} (see also \cite{Suyama:2010uj}).

If inflation is driven by more than one degree of freedom, the end of inflation can differ significantly: instead of a sudden end caused by the simultaneous decay of all fields, a stretched out decay phase is possible \cite{Battefeld:2008py,Battefeld:2008ur,Battefeld:2008qg}, and in some cases unavoidable \cite{Becker:2005sg,Ashoorioon:2006wc,Ashoorioon:2008qr}, and the feasibility of (p)reheating can change drastically \cite{Battefeld:2008bu,Battefeld:2008rd,Battefeld:2009xw,Braden:2010wd}. Staggered inflation \cite{Battefeld:2008py,Battefeld:2008ur,Battefeld:2008qg} is not a new type of inflation, but a collective term coined for models that contain such decaying fields during inflation. If all fields decay in a few e-folds, as in \cite{Becker:2005sg}, the preheating phase is merely extended, without strong observational consequences, but if they decay throughout the last sixty e-folds, additional signatures in the correlation functions of fluctuations in the Cosmic Microwave Background (CMB) are possible \cite{Battefeld:2008py,Battefeld:2008ur,Battefeld:2008qg,Battefeld:2010rf} (see also \cite{Barnaby:2010ke,Barnaby:2009dd,Barnaby:2009mc} for additional signatures caused by particle production during inflation, as present in trapped inflation \cite{Kofman:2004yc,Green:2009ds,Battefeld:2010sw}).

We use the term ``decay'' loosely to indicate the partial or full transfer  of an inflaton's energy (potential and/or kinetic) to an additional component of the energy momentum tensor, such as radiation. A concrete example is the annihilation of a brane/antibrane pair in the extension of the KKLMMT proposal in \cite{Battefeld:2010rf}; whenever a brane comes in close contact to an anti-brane, they annihilate and produce closed string modes that redshift like radiation. Other examples include inflation driven by tachyons \cite{Piao:2002vf,Majumdar:2003kd}, by multiple M5-branes \cite{Becker:2005sg,Ashoorioon:2006wc,Ashoorioon:2008qr} or inflation on the landscape \cite{Battefeld:2008qg}, among others. 

If many fields decay in any given Hubble time, an analytic formalism, based on coarse graining, was developed in \cite{Battefeld:2008py,Battefeld:2008qg}; the aim of this study was to retain some of the effects caused by the decaying fields, such as contributions to the power-spectrum generated by the additional decrease of the energy that drives inflation, due to the fields' decay as opposed to slow roll. As a consequence, contributions proportional to $\Gamma/H$, where $\Gamma$ is the decay rate and $H$ the Hubble parameter, appear alongside the common slow roll parameters in observables such as the scalar spectral index or the tensor to scalar ratio; furthermore, the new contributions may even be the dominant ones \cite{Battefeld:2008qg}. 

However, any effect due to the sudden decay of an individual field is not retained. To ameliorate this shortcoming, a single decay was discussed in detail in a concrete  setup (an extension of the KKLMMT proposal) in \cite{Battefeld:2010rf} and in a related DBI-inflation setup \cite{Firouzjahi:2010ga}. There it was shown that a jump in the equation of state parameter results after the decay.  This discontinuity in turn causes mode mixing at the perturbed level \footnote{Mode-mixing means that the second independent solution of the Sasaki-Mukhanov variable, which has zero amplitude if Bunch-Davies vacuum initial conditions are imposed, acquires a non-zero amplitude.}. In \cite{Battefeld:2010rf} the matching conditions for perturbations were derived in the sudden decay approximation, neglecting perturbations in radiation. 
A ringing pattern on top of the nearly scale invariant power-spectrum was found for sub-horizon modes \cite{Battefeld:2010rf} (see also \cite{Nakashima:2010sa} for the same effect induced not by decays, but by a varying speed of sound) and super-horizon modes showed small corrections. Extending the model of \cite{Battefeld:2010rf} further, to allow for many decays, should yield a power-spectrum that converges to the one of \cite{Battefeld:2008py,Battefeld:2008qg} even for sub-horizon modes, as a superposition washes out any signals of the individual decays. 

To this end, we develop the numerical tools needed for such a comparison in this paper. We apply them to a much simpler setup motivated by inflation on the landscape \cite{Battefeld:2008qg}, that served as a case study for the formalism in \cite{Battefeld:2008py} and enables us to compare results. We plan to use the code developed for this study in the more complicated, yet more realistic model of \cite{Battefeld:2010rf} in the near future.

Besides developing and testing the numerical code, we show how the analytic treatment of staggered inflation in \cite{Battefeld:2008py,Battefeld:2008qg} is recovered in the large $\mathcal{N}$-limit, and argue that this formalism serves as an excellent approximation if more than $\sim 40$ fields decay in any given Hubble time.

The outline of this paper is as follows: in Sec.~\ref{sec:bgrmodel} we specify the model before reviewing the analytic results of \cite{Battefeld:2008py,Battefeld:2008qg} in Sec.~\ref{sec:largeNlimit}. The matching conditions derived in \cite{Battefeld:2010rf}, which are needed for the numerical treatment, are provided in Sec.~\ref{sec:match}. We then show in Sec.~\ref{sec:numerics} how the relatively simple results of \cite{Battefeld:2008qg} are recovered numerically in the large $\mathcal{N}$-limit, while we recover signatures of individual decays if only a few fields are present. Details about the numerical code can be found in the Appendix.

If not stated otherwise we set the reduced Planck mass equal to one, $m_{pl}^{-2}=8\pi G\equiv 1$ and use ``$\simeq$'' to denote equality to leading order in small parameters.

\section{Staggered Multi-Field Inflation}
\subsection{The Background Model \label{sec:bgrmodel}}
Consider $\mathcal{N}$ uncoupled, scalar fields $\phi_I$, $I=1\dots \mathcal{N}$ with canonical kinetic terms and linear potentials
\begin{eqnarray}
V_I=V_0^I-c_I\phi_I\,. \label{potential1}
\end{eqnarray}
Such a potential is motivated by expanding a general potential on the string-landscape for fields that reside on flat stretches \cite{Battefeld:2010rf}; fields located on steep slopes evolve faster and become dynamically irrelevant for the evolution of the universe. Hence, we are interested in a narrow distribution of the $c_I$ and fields with comparable potential energies. To keep the model as simple as possible, we consider $c_I=c_J\equiv c$ and $V_0^I\equiv V_0$ for all $I,J$ (see \cite{Battefeld:2010rf} for more general setups).

 Such an expansion is expected to be valid for small ranges of field values only, so we use (\ref{potential1}) only up until some maximal field value $\phi_I=\phi_{end}\ll 1$; 
if a field encounters $\phi_{end}$, we assume that its potential energy is converted into an additional component of the energy momentum tensor, i.e.~radiation $\rho_r$. Thereafter, the field may be stabilized or continue to roll freely. In either case the field quickly becomes irrelevant for the evolution of the universe.  We choose to set the potential to zero for $\phi_I>\phi_{end}$ and let fields roll freely for larger field values, always keeping $\phi_I$ and $\dot{\phi}_I$ continuous through time\footnote{Since the kinetic energy is negligible compared to the potential one, we do not expect any significant changes if part or all of the kinetic energy were also infused into radiation.}. Since we would like the previous stretches to be flat, we demand
\begin{eqnarray}
 \frac{c_I\phi_{end}}{V_0}\ll 1\,. \label{tildevarepsilon}
\end{eqnarray}
Depending on initial conditions, some fields encounter $\phi_{end}$ earlier than others, causing ``decaying'' or ``dropping'' fields to enter the freely rolling phase in a staggered fashion. As we have no knowledge about the concrete initial conditions, we distribute the initial field values randomly over the interval $(0,\phi_{end})$, from which they evolve according to their Klein-Gordon equations
\begin{eqnarray}
\ddot{\phi}_I+3H\dot{\phi}_I=-\frac{\partial V_I}{\partial \phi_I}\,,
\end{eqnarray}
with $V$ from (\ref{potential1}) for $\phi_I<\phi_{end}$ and $V=0$ otherwise.  Radiation is produced whenever a field encounters $\phi_{end}$, taking over the remaining potential energy and redshifting as
\begin{eqnarray}
\dot{\rho}_r=-4H\rho_r
\end{eqnarray}
thereafter.

\subsection{Analytic Results in the Large $\mathcal{N}$ Limit \label{sec:largeNlimit}}
If many fields drop out during any given Hubble time, one can describe the system analytically \cite{Battefeld:2008py,Battefeld:2008qg} by promoting the number of fields to a continuous function that decreases in time according to the decay rate
\begin{eqnarray}
\Gamma \equiv -\frac{\dot{\mathcal{N}}}{\mathcal{N}}>0\,.
\end{eqnarray}
This rate is not a new free function but set by the initial values and subsequent evolution of the fields. Once $\Gamma\sim H$, inflation ends because all remaining fields decay in a single Hubble time. Hence, inflation driven by many fields can take place as long as $\Gamma\ll H$, that is as long as \cite{Battefeld:2008py}
\begin{eqnarray}
\varepsilon_{\mathcal{N}}\equiv \frac{\Gamma}{2H} \ll 1 \,,
\end{eqnarray}
where we included a $2$ in the definition of $\varepsilon_{\mathcal{N}}$ to simplify expressions below.

If inflation is of the slow roll type, that is if the slow roll parameters $\varepsilon_I\equiv (\partial V_I/\partial \varphi_I)^2/(2W^2)$, $\varepsilon\equiv\sum_I\varepsilon_I$ as well as $\eta_I\equiv (\partial^2 V_I/\partial \varphi_I^2)/W$ and $\eta\equiv \sum_I \eta_I$ are small (see \cite{Easther:2005nh} for details on the Hubble slow roll expansion for multi-field inflation),
one can show \cite{Battefeld:2008py} that the combined energy of the fields and the one of radiation obey
\begin{eqnarray}
\dot{\rho}_{\mbox{\tiny inf}} & \simeq & -2H(\varepsilon_{\mathcal{N}}+\varepsilon)\rho_{\mbox{\tiny inf}}\,,\\
\dot{\rho}_r& \simeq & -2H \left(\frac{3}{2}(1+w_r)\rho_r -\varepsilon_{\mathcal{N}}\rho_{\mbox{\tiny inf}}\right)\simeq 2H (\varepsilon_{\mathcal{N}}-\bar{\varepsilon})\rho_{\mbox{\tiny inf}}\,,
\end{eqnarray}
to first order in small parameters. Here $w_r=p_r/\rho_r=1/3$ and
\begin{eqnarray}
\bar{\varepsilon}\equiv \frac{3}{2}(1+w_r)\frac{\rho_r}{\rho_r+\rho_{\mbox{\tiny inf}}} \simeq \varepsilon_{\mathcal{N}}\,.\label{barepsilon}
\end{eqnarray}
In the limit of many decays during any given Hubble time, radiation approaches a scaling solution where the energy loss due to redshifting is compensated by the infusion of energy from the inflaton sector, $\rho_r\simeq \varepsilon_{\mathcal{N}}2\rho_{I}/(3+3w_r)$.
 Further, the Hubble slow evolution parameter becomes \cite{Battefeld:2008py}
 \begin{eqnarray}
-\frac{\dot{H}}{H^2}\simeq \varepsilon+\bar{\varepsilon}\,.\label{Hubbleslow}
 \end{eqnarray}
 
The analysis of adiabatic perturbations is straightforward, albeit tedious, and one can show that in certain cases (for example in the model we investigate in this paper) the effects of isocurvature perturbations are negligible \cite{Battefeld:2008py}. The scalar power-spectrum of the curvature perturbation on uniform density surfaces $\zeta_k$ becomes \cite{Battefeld:2008py,Battefeld:2008qg}
\begin{eqnarray}
\mathcal{P}_{\zeta}
\simeq \frac{H^2}{8\pi^2 m_{pl}^2(\varepsilon \gamma^2 + \bar{\varepsilon})}\,, \label{powerzeta}
\end{eqnarray}
where we reinstated the reduced Planck mass and $\gamma$ is a parameter of order one set by the background evolution \cite{Battefeld:2008py}; since we are primarily interested in situations where the staggered inflation effects dominates over slow roll effects, $\varepsilon,\eta\ll \bar{\varepsilon}$, we do not need to compute $\gamma$ and the scalar spectral index in \cite{Battefeld:2008py,Battefeld:2008qg} simplifies to 
\begin{eqnarray}
n_s-1&=&\frac{d\,\ln \mathcal{P}_{\zeta}}{d\, \ln k}\\
&\simeq & (\delta-3)  \bar{\varepsilon}\,, \label{scalarspectralindex}
\end{eqnarray}
where 
\begin{eqnarray}
\delta \equiv \frac{\dot{\Gamma}H}{\Gamma \dot{H}}\,.
\end{eqnarray}
Similarly, the tensor power-spectrum can be computed to
\begin{eqnarray}
\mathcal{P}_{T}&\equiv &2\frac{4\pi k^3}{(2\pi)^3}\left|h_k^2\right|\\
&\simeq &\frac{2}{\pi^2}\frac{H^2}{m_{pl}^2}\,, \label{powertensor}
\end{eqnarray}
with the tensor spectral index
\begin{eqnarray}
n_{T}&\equiv& \frac{d\, \ln \mathcal{P}_T}{d\, \ln k}\\
&\simeq &-2\bar{\varepsilon} \,. \label{tensorspectralindex}
\end{eqnarray}
and the tensor to scalar ratio
\begin{eqnarray}
r&\equiv &\frac{\mathcal{P}_T}{\mathcal{P}_\zeta}\\
 &\simeq &16 \bar{\varepsilon}\,. \label{tensortoscalarratio}
\end{eqnarray}

For the linear potential in (\ref{potential1}), we get
\begin{eqnarray}
\Gamma\simeq \frac{c}{\varphi_{end}\sqrt{3V_0\mathcal{N}}}\,,
\end{eqnarray}
so that 
\begin{eqnarray}
\bar{\varepsilon}\simeq \varepsilon_{\mathcal{N}}=\frac{\Gamma}{2H}\simeq\frac{c}{2\varphi_{end}V_0\mathcal{N}}\,. \label{barepsilonlinear}
\end{eqnarray}
and
\begin{eqnarray}
\delta\simeq -1\,.
\end{eqnarray} 
The number of e-folds $N\approx 60$ becomes 
\begin{eqnarray}
N&=&\int_{ini}^{end} H\, dt
\simeq\frac{V_0\varphi_{end}\mathcal{N}}{2c} \,, \label{efoldslinear}
\end{eqnarray}
which allows us to express $\bar{\varepsilon}$ in terms of $N$,
\begin{eqnarray}
\bar{\varepsilon}\simeq \frac{1}{4N}\,.
\end{eqnarray}
The resulting scalar spectral index (\ref{scalarspectralindex}), tensor spectral index (\ref{tensorspectralindex}) and tensor to scalar ratio (\ref{tensortoscalarratio}) are
\begin{eqnarray}
n_s-1&\simeq&-\frac{1}{N}\,,\label{nslinear}\\
n_T&\simeq&-\frac{1}{2N}\,,\\
r&\simeq&\frac{4}{N}\,. \label{rlinear}
\end{eqnarray}
These expressions are valid if the slow roll contributions are subdominant and many fields drop out in any given Hubble time so that the coarse grained analytic treatment is justified. All signatures due to the sudden changes in the equation of state parameter whenever a field drops out are not retained.

\subsection{Model Parameters \label{modelparameters}}
In the next section we aim to test these analytic predictions numerically, without performing a coarse graining. To this end, we need to specify the model parameters. To avoid confusion, we restore the reduced Planck mass $m_{pl}=(8\pi G)^{-1/2}$ in this section. First, we want to compute observables around sixty e-folds before the end of inflation when modes that are observable in the CMB crossed the Hubble horizon,
\begin{eqnarray}
N\equiv 60\,.
\end{eqnarray}
We will vary the number of fields $\mathcal{N}$, but are particularly interested in the limit $\mathcal{N}\gg 1$. Furthermore, we would like to impose $\varepsilon, c_I\phi_{end}/V_0 \ll \bar{\varepsilon}$; to be concrete we set
\begin{eqnarray}
\varepsilon=\frac{\mathcal{N}}{2m_{pl}^2}\left(\frac{c}{\sum_I(V_0-c\varphi_{I})}\right)^2\simeq \frac{\mathcal{N}}{2 m_{pl}^2}\left(\frac{c}{V_0}\right)^2\equiv \frac{1}{2}\bar{\varepsilon}^2\,,
\end{eqnarray}
so that 
\begin{eqnarray}
c=\frac{V_0\sqrt{\mathcal{N}}}{4N m_{pl}}\,.
\end{eqnarray}
The inflationary scale follows from (\ref{powerzeta}) with the COBE normalization $\mathcal{P}_{\zeta}\approx \mathcal{P}_{\mathcal{R}}\approx 2.4\times 10^{-9}$, so that
\begin{eqnarray}
V_0=3\frac{8\pi^2}{4\mathcal{N}\bar{N}}2.4\times 10^{-9}m_{pl}^4\,,
\end{eqnarray}
where we used $\varepsilon\ll \bar{\varepsilon}$. The critical field value follows from (\ref{efoldslinear}) as
\begin{eqnarray}
\varphi_{end}=\frac{1}{2\sqrt{\mathcal{N}}}m_{pl}\,.
\end{eqnarray}
One may check that $c_I\phi_{end}/V_0=\bar{\varepsilon}/(2\sqrt{\mathcal{N}})\ll \bar{\varepsilon}$, as desired.

\section{Perturbations}
We now go beyond the analytic approximations in \cite{Battefeld:2008py,Battefeld:2008qg}, such as the large $\mathcal{N}$-limit and the slow roll approximation, and investigate the evolution of scalar perturbations numerically. 

The line element to linear order in scalar perturbations and without fixing a gauge is
\begin{eqnarray}
ds^2 = -(1 + 2A)dt^2 + 2aB_{,i}dx^i dt + a^2 [(1 - 2\psi)\delta_{ij}
+ 2E_{,ij}]  dx^i dx^j\,.
\end{eqnarray}
Metric degrees of freedom couple to the perturbations of the scalar fields $\delta\phi_I$ in their equations of motion \cite{Taruya:1997iv, Gordon:2000hv}
\begin{eqnarray}
\ddot{\delta\phi}_I+3H\dot{\delta\phi}_I+\frac{k^2}{a^2}\delta\phi_I+\sum_I
V_{,\phi_I\phi_J}\delta\phi_J=-2V_{,\phi_I}A+\dot{\phi_I}\left[\dot{A}+3\dot{\psi}+\frac{k^2}{a^2}\left(a^2\dot{E}-aB\right)\right]\,,
\end{eqnarray}
with $I=1,\dots,\mathcal{N}$. Two metric degrees of freedom can be eliminated by the choice of gauge.  Utilizing this choice, we use two gauge-invariant perturbations, i.e.~the two Bardeen potentials
\cite{Mukhanov:1990me}
\begin{eqnarray}
\Phi &=& A+\left(aB-a^2\dot{E}\right)^{.}\,, \label{Bardeen1}\\
\Psi &=&\psi-H \left(aB-a^2\dot{E}\right)\,.\label{Bardeen2}
\end{eqnarray}
Since anisotropic stress is absent in our setup, we have $\Phi=\Psi$. It is also useful to introduce gauge invariant field perturbations, for instance the
Sasaki-Mukhanov variables,
\begin{eqnarray}
Q_I=\delta \phi_I+\frac{\dot{\phi}_I}{H}\psi\,,
\end{eqnarray}
where the spatially flat gauge is defined by the condition $\psi=0$. The equations of motion for the $Q_I$ are \cite{Taruya:1997iv,Gordon:2000hv} 
\begin{eqnarray}
\label{Q-eq}
0&=&\ddot{Q}_I+3H\dot{Q}_I+\frac{k^2}{a^2}Q_I+\sum_J\left(V_{,\phi_I\phi_J}-\frac{1}{m_{pl}^2a^3}\left(\frac{a^3}{H}\dot{\phi}_I\dot{\phi}_J\right)^{\!.}\right)Q_J\,
\end{eqnarray}
in the absence of perturbations in additional degrees of freedom. Even though we keep track of radiation at the background level, we ignore perturbations in $\rho_r$. 

The Sasaki-Mukhanov variables have the advantage that their evolution is decoupled from the metric degrees of freedom. We assume that the $Q_{I}$ are in the  Bunch Davies vacuum at the onset of inflation, for $-k\tau\rightarrow \infty$, that is 
\begin{eqnarray}
Q_I(t_t)\equiv \frac{u_I(\tau_i)}{a(\tau_i)}= \frac{e^{-ik\tau}}{a(\tau_i)\sqrt{2k}} {\bf e}_I\,.
\label{vacinit}
\end{eqnarray}
where the ${\bf e}_I$ are independent unit Gaussian random fields with
\ba
 < {\bf e}_I > =0 \, , \quad < {\bf e}_I({\bf k}) \, {\bf e}_J({\bf k'}) >
 = \delta_{IJ} \delta^3 ( {\bf k} - {\bf k'} ) \, .
 \ea
Since our ultimate goal is to numerically solve the equations of motion (\ref{Q-eq}), we want to use analytic solutions up until the wavelength of a mode approaches the horizon size. However, whenever a field drops out, the analytic solutions do not conform to the vacuum solution any more, but carry an admixture of the second independent solution of (\ref{Q-eq}),
\begin{eqnarray}
Q_I(t_t)&=& \frac{1}{a}\frac{\sqrt{-\pi\tau}}{2}e^{i\pi(\mu+1/2)/2}\left(\alpha H_\mu^{(1)}(-k\tau)+\beta H_{\mu}^{(2)}(-k\tau)\right){\bf e}_I\,,\label{analyticolQHankel}\\
&\approx& \frac{1}{a\sqrt{2k}}\left(\alpha_I e^{-ik\tau}+\beta_I e^{i k \tau}\right){\bf e}_I\,, \label{analyticolQ}
\end{eqnarray}
where we expand the Hankel functions for large arguments and used $\mu=3/2+\mathcal{O(\varepsilon,\bar{\varepsilon},\eta)}\approx 3/2$ during inflation. The Bogoliubov coefficients $\alpha$ and $\beta$ need to be computed according to the matching conditions of the next section whenever a field drops. 

\subsection{Matching Conditions \label{sec:match}}
When fields encounter the sharp drop in the potential, the equation of state parameter makes a jump due to the creation of radiation. We follow closely \cite{Battefeld:2010rf} where the matching conditions of perturbations were derived in a related setup \footnote{The simpler matching conditions in \cite{Firouzjahi:2010ga} lead to jumps in the extrinsic curvature and/or induced metric on the transition hypersurface in contradiction to \cite{Deruelle:1995kd,Martin:1997zd}, but might still provide a viable approximation for super-horizon modes.}. 
Based on the continuity of the induced metric and the extrinsic curvatures on the hyper-surface at which the equation of state parameter jumps, one can derive the matching conditions for the Sasaki-Mukhanov variables \cite{Israel:1966rt, Deruelle:1995kd, Martin:1997zd}. The transitions, which we model as instantaneous events\footnote{In more realistic scenarios, our treatment remains reliable for $k\ll (\Delta t)^{-1}$ where $\Delta t$ is the time-scale of the transition.}, occur at  well defined values of the inflatons. As a result, the Bardeen potential and the comoving curvature perturbation are both continuous\footnote{We ignore terms of order $\mathcal{O}(k^2)$, since we are interested in the effects of inflaton decays during the last sixty e-folds onto super-horizon modes.} \cite{Zaballa:2009xb,Lyth:2005ze,Zaballa:2006kh}
\begin{eqnarray}
\label{match1}
[ \Phi ]_\pm =0 \, , \quad 
\label{second-b.c.}[\mathcal{R}]_{\pm}=0\, ,
\end{eqnarray}
where
\begin{eqnarray}
\label{defR}
\mathcal{R}&=&\Phi+
\frac{2}{3(1+w)}\left(\frac{\Phi^\prime}{\mathcal{H}}+\Phi\right)
\,,
\end{eqnarray}
The subscripts $-$ and $+$ refer to the value of the quantity in brackets before and after the field drops out, respectively.  At the background level, the scale-factor, Hubble parameter $\mathcal{H}= a^\prime/a$ and velocities of the fields are continuous 
\begin{eqnarray}
{[ a ]_\pm=0} \;\;\;\; ,\;\;\;\; {[ \mathcal{H}]_\pm=0} \, \;\;\;\; ,\;\;\;\;  {[ \phi_I^\prime]_\pm=0}.
\end{eqnarray}
The comoving curvature perturbation, $\mathcal{R}$, is related to the Sasaki-Mukhanov variables via
\cite{Gordon:2000hv}
\begin{eqnarray}
\mathcal{R}
&=& \frac{1}{3 (1+ w) m_{pl}^2 \mathcal{H}}  \sum_I \phi_I^\prime Q_I
\,. \label{defR2}
\end{eqnarray}
If no additional contributions to the energy momentum tensor are present at the perturbed level, we still need to express $\Phi$ in terms of the $Q_I$ to derive their matching conditions; to this end, we need the perturbed Einstein equations \cite{Mukhanov:1990me}
 \begin{eqnarray}
 -3\mathcal{H}(\mathcal{H}\Phi+\Psi^\prime)-k^2\Psi&=&\frac{1}{2m_{pl}^2}a^2\delta T_0^{(gi)\,0}\,,\label{PEEQ1}\\
 (\mathcal{H}\Phi+\Psi^\prime)_{,i}&=&\frac{1}{2m_{pl}^2}a^2\delta T_i^{(gi)\,0}\,, \label{PEEQ2}
 \end{eqnarray}
 with the gauge invariant perturbations of the energy momentum tensor
 \begin{eqnarray}
 \delta T_0^{(gi)\,0}&=&\frac{1}{a^2}\sum_I\left(-\phi_I^{\prime 2}\Phi+\phi_I^\prime\delta\phi_I^{(gi)\,\prime}+V_{,\phi_I}a^2\delta\phi_{I}^{(gi)} \right)\,,\\
 \delta T_i^{(gi)\,0}&=&\frac{1}{a^2}\sum_I\phi_I^\prime\delta \phi_{I,i}^{(gi)}\,,
 \end{eqnarray}
 and the gauge invariant field perturbation
 \begin{eqnarray}
 \delta \phi_{I}^{(gi)}=\delta\phi_I+\phi_I^\prime(B-E^\prime)\,.
 \end{eqnarray}
After some algebra \cite{Battefeld:2010rf} we arrive at
 \begin{eqnarray}
 \label{defPhi}
 -k^2\Phi&=&
 \frac{1}{2 m_{pl}^2}\sum_I\left( Q_I^\prime\phi_I^\prime+Q_I\left(a^2V_{,\phi_I}+\frac{3}{2}(1- w) \phi_I^{\prime}\mathcal{H}\right) \right)\,.
 \end{eqnarray}
In order to guarantee the continuity of $\mathcal{R}$ and $\Phi$, we impose matching conditions for all elements of the sums in (\ref{defR2}) and (\ref{defPhi}) individually,
 \ba
 \label{match2}
 \left[ \frac{Q_I}{1+w}\right]_\pm =0 \,,\\
 \left[  Q_I' \phi_I^\prime+Q_I\left(a^2V_{,\phi_I}+\frac{3}{2}(1-w)\phi_I^\prime\mathcal{H}\right)\right]_\pm =0 \,.
 \ea
 For ease of notation we define
 \begin{eqnarray}
 A_I\equiv a^2\frac{V_{,\phi_I}}{\phi_I^\prime \mathcal{H}}+\frac{3}{2}(1-w)\,. \label{shorthandA}
 \end{eqnarray}
(\ref{match2}) leads to an enhancement of $Q_I$ on all scales each time an inflaton encounters a critical value. The magnitude of the jump is determined by the ratio 
\begin{eqnarray}
B\equiv \frac{1+w_+}{1+w_-}\,.
\end{eqnarray}
If $\varepsilon \ll \bar{\varepsilon}$, as in Sec.\ref{modelparameters}, the equation of state parameter becomes to leading order in $\bar{\varepsilon}$
\begin{eqnarray}
1+w\simeq \frac{2}{3}\bar{\varepsilon}= 2\frac{\rho_r}{\rho_r+\rho_{\mbox{\tiny inf}}}\,, \label{1pluswsr}
\end{eqnarray}
so that
\begin{eqnarray}
B\simeq \frac{\rho_r^+}{\rho_r^-}\approx \frac{1}{e^{4\Delta N}}\,,
\end{eqnarray}
where $\Delta N\sim \mathcal{O}(N/\mathcal{N})$ is the number of e-folds between successive drop of fields. Thus, if many fields drop out in any given Hubble time, $\mathcal{N}\gg N$, the jumps become exceedingly small. In \cite{Battefeld:2010rf}, it was found in a related model that the deviation of $B$ from one also determines the amplitude of additional features in the power-spectrum. Based on this, we expect a requirement of $N/\mathcal{N} \ll 1$, that is at least $\mathcal{N}\sim \mathcal{O}(10^{3})$ fields, to remain observationally viable.

Let's turn our attention to the second matching condition, (\ref{match2}). If the system is in a slow roll regime, the speed $Q^\prime_I$ is continuous to zeroth order in small parameters for all but the one field whose potential energy has been infused into radiation. To see this, we insert the slow roll equation of motion, $3\mathcal{H}\phi^\prime=-a^2V_{,\phi_I}$, into (\ref{shorthandA}) and work to zeroth order in slow roll to arrive at
\begin{eqnarray}
A_I\simeq \frac{3}{2}(1+w)\,.
\end{eqnarray}
Thus, using (\ref{1pluswsr}) and working to zeroth order in $\bar{\varepsilon}$ and the slow roll parameters, one may use  
\begin{eqnarray}
 \left[  Q_I' \right]_\pm \simeq 0 \,,
\end{eqnarray}
instead of (\ref{match1}). This matching condition is a good approximation for the fields responsible for reheating the universe after inflation, which do not encounter $\phi_{end}$ until the end of inflation\footnote{We use the full matching conditions in our numerical code.}. These fields determine the scalar power-spectrum and are thus the ones of primary interest to us.
 
\subsubsection{Matching the Bogoliubov Coefficients}
As long as $-k\tau\gg 1$, we rely on the analytic approximation for the $Q_I$ in (\ref{analyticolQ}). Initially, $\alpha_I=1$ and $\beta_I=0$, but $\beta_I\neq 0$ as soon as the first field encounters the step in the potential. At each of those instances we have to match  the analytic solution in (\ref{analyticolQ}) according to (\ref{match2}) and (\ref{match1}), yielding
\begin{eqnarray}
\nonumber \alpha^+_I&=&\alpha^-_I\frac{1}{2}\left(1+\frac{1+w_+}{1+w_-}+i\frac{\mathcal{H}}{k}\left(\frac{1+w_+}{1+w_-}(1-A^+_I)-(1-A^-_I)\right)\right)\\
 &&-\beta^-_I\frac{1}{2}\left(1-\frac{1+w_+}{1+w_-}-i\frac{\mathcal{H}}{k}\left(\frac{1+w_+}{1+w_-}(1-A^+_I)-(1-A^-_I)\right)\right)e^{-2ik/\mathcal{H}}\,,\label{alpha}\\
\nonumber \beta^+_I&=&-\alpha^-_I\frac{1}{2}\left(1-\frac{1+w_+}{1+w_-}+i\frac{\mathcal{H}}{k}\left(\frac{1+w_+}{1+w_-}(1-A^+_I)-(1-A^-_I)\right)\right)e^{2ik/\mathcal{H}}\\
&&+\beta^-_I\frac{1}{2}\left(1+\frac{1+w_+}{1+w_-}-i\frac{\mathcal{H}}{k}\left(\frac{1+w_+}{1+w_-}(1-A^+_I)-(1-A^-_I)\right)\right)\,,\label{beta}
\end{eqnarray}
with $A$ from (\ref{shorthandA}) and we used $-\tau\simeq \mathcal{H}^{-1}$. One can check that
\begin{eqnarray}
 \left|\alpha_I^+\right|^2-\left|\beta_I^+\right|^2=\frac{1+w_+}{1+w_-}\left(\left|\alpha_I^-\right|^2-\left|\beta_I^-\right|^2\right)\,,
\end{eqnarray}
as in \cite{Battefeld:2010rf}. Note that all modes need to be renormalized\footnote{The source of $\left|\alpha_I^+\right|^2-\left|\beta_I^+\right|^2\neq 1$ is our neglection of perturbations in radiation.} by $1/\sqrt{B}$  after the matching to guarantee $\left|\alpha_I^+\right|^2-\left|\beta_I^+\right|^2=1$.

There is, however, another subtlety related to the sudden decay approximation $\Delta \tau_d\rightarrow 0$: in this limit all modes within the horizon are affected by the transition, leading to unsuppressed oscillatory corrections on top of the power-spectrum for all $-k\tau\gg 1$ \cite{Battefeld:2010rf}. However, in a realistic setting the infusion of potential energy from an inflationary field into radiation needs time. We do not expect modes with wavelength much smaller than this time-scale ($k\gg (\Delta \tau_d)^{-1}$) to be affected. This in turn should lead to damping of the oscillations on top of the power-spectrum. Since we are working from a phenomenological viewpoint, we treat $\Delta \tau_d$ as a free parameter\footnote{In the extension of the KKLMMT model in \cite{Battefeld:2010rf}, this time-scale is set by the brane decay time.}. 

To incorporate this effect into the matching conditions (\ref{alpha}) and (\ref{beta}) we define a simple smooth window function
\begin{eqnarray}
F(k)\equiv\frac{1}{2}\left(1+\tanh\left[\tilde{c} \left(\log\left(\frac{k}{k_{t}}\right)-1\right)\right]\right)\,. \label{windowfunction}
\end{eqnarray}
The value of $\tilde{c}$ determines the sharpness of the window function in k-space and $k_{t}$ indicates the limiting wave-number above which modes are unaffected by the decay. 

This window function can be used to modify
the matching conditions to
\begin{eqnarray}
\alpha^+_{I}&=&\sqrt{C_I}\left(\sqrt{F}+\sqrt{\frac{1-F}{B}} \bar{\alpha}^+_{I}\label{alphanew}\right)\,,\\
\beta^+_{I}&=&\sqrt{C_I}\left(\sqrt{\frac{1-F}{B}} \bar{\beta}^+_{I}\,,\label{betanew}\right)\,,
\end{eqnarray}
where we denote with $\bar{\alpha}^+_{I}$ and $\bar{\beta}^+_{I}$ the right hand side of (\ref{alpha}) and (\ref{beta}) respectively and we define
\begin{eqnarray}
C_I\equiv \frac{1}{1+\sqrt{F(1-F)/B} \left(\bar{\alpha}^+_{I}+\bar{\alpha}^{+*}_{I}\right)}\,.
\end{eqnarray}
The $C_I$ and the factors of $B$ guarantee the proper normalization of the Bogoliubov coefficients, $\left|\alpha_I^+\right|^2-\left|\beta_I^+\right|^2=1$. Note that the $C_I$ only differ from one in a region around $k_t$, whose width is set by the sharpness of the window function, $\tilde{c}$. We won't make use of the smooth window-function in our code, but use a sharp theta-function for simplicity.

\subsection{The Power-spectrum}
The power-spectrum of the curvature perturbation\footnote{$\mathcal{R}_k$ coincides with $\zeta_k$ on large scales, the difference being terms of order $k^2$.} is defined as (see \cite{Bassett:2005xm} for a review)
\begin{eqnarray}
\delta^3({\bf k}-{\bf k}^\prime)\mathcal{P}_{\mathcal{R}}=\frac{4\pi
k^3}{(2\pi)^3}<\mathcal{R}({\bf k}^\prime)^*\mathcal{R}({\bf k})> \label{powerspectrum}
\end{eqnarray}
and the scalar spectral index is given by
\begin{eqnarray}
n_s-1\equiv d \ln
\mathcal{P}_{\mathcal{R}}/ d \ln k\,.
\end{eqnarray}

\section{Numerical Results and Comparison to Analytics \label{sec:numerics}}
We solve the staggered inflation model of section \ref{sec:bgrmodel} numerically with the model parameters of section \ref{modelparameters} and a varying number of fields $\mathcal{N}=1\dots 2500$. Perturbations are treated analytically deep inside the horizon, according to (\ref{analyticolQ}), with or without adjusted Bogoliubov coefficients according to (\ref{alphanew}) and (\ref{betanew}) whenever a field drops/decays. Not imposing the matching corresponds to a sharp window function at $k_{\mbox{\tiny num}}=e^{4} H_{ini}$; once a mode approaches the horizon, we solve (\ref{Q-eq}) numerically, together with the matching conditions in (\ref{match2}) and (\ref{match1}). More details on the code can be found in appendix \ref{sec:code}. Towards the end of inflation, we evaluate the power-spectrum in (\ref{powerspectrum}), which is dominated by perturbations in the fields that drove inflation the longest, and read off the spectral index.

How many fields need to decay in any given Hubble time for the large $\mathcal{N}$ limit of
Sec.~\ref{sec:largeNlimit} \cite{Battefeld:2008py,Battefeld:2008qg} to be a good approximation? To get a simple order of magnitude estimate, consider $\bar{\epsilon}=2\rho_r/(\rho_r+\rho_{inf})$ in (\ref{barepsilon}) and assume that around $60/\mathcal{N}$ fields drop out in a Hubble time. $\bar{\epsilon}\sim \mathcal{O}(10^{-2})$ needs to be smooth to at least one part in ten in order to describe properly the deviation of the power-spectrum from scale invariance, (\ref{nslinear}). Lets focus on the relevant regime around sixty e-folds before the end of inflation when $\mathcal{N}$ is large. Since $\Delta \rho_r=\rho_{inf}/\mathcal{N}$ we get from $\Delta \bar{\epsilon}/\bar{\epsilon}\sim 2/(\mathcal{N}10^{-2})< 10^{-1}$ a lower bound of $\mathcal{N}\geq 2000$. Another way of getting an estimate is to consider how much $\rho_r$ redshifts until the next field drops out: since $\bar{\epsilon}\sim 10^{-2} (a_0/a(t))^4$ and $a(t)/a_0\approx \exp(H\Delta t)\approx 1+H\Delta t$ we get by using $\Delta t\sim 60/\mathcal{N}$ a lower bound of $\mathcal{N}\geq 2400$. Thus we need around
\begin{eqnarray}
\mathcal{N}\gtrsim \mathcal{O}(10^3)\,,
\end{eqnarray}
fields initially.

\begin{figure}[htb]

\begin{tabular}{cc}
\subfloat[]
{\label{fig:1a}
\includegraphics[width= 0.35\textwidth]{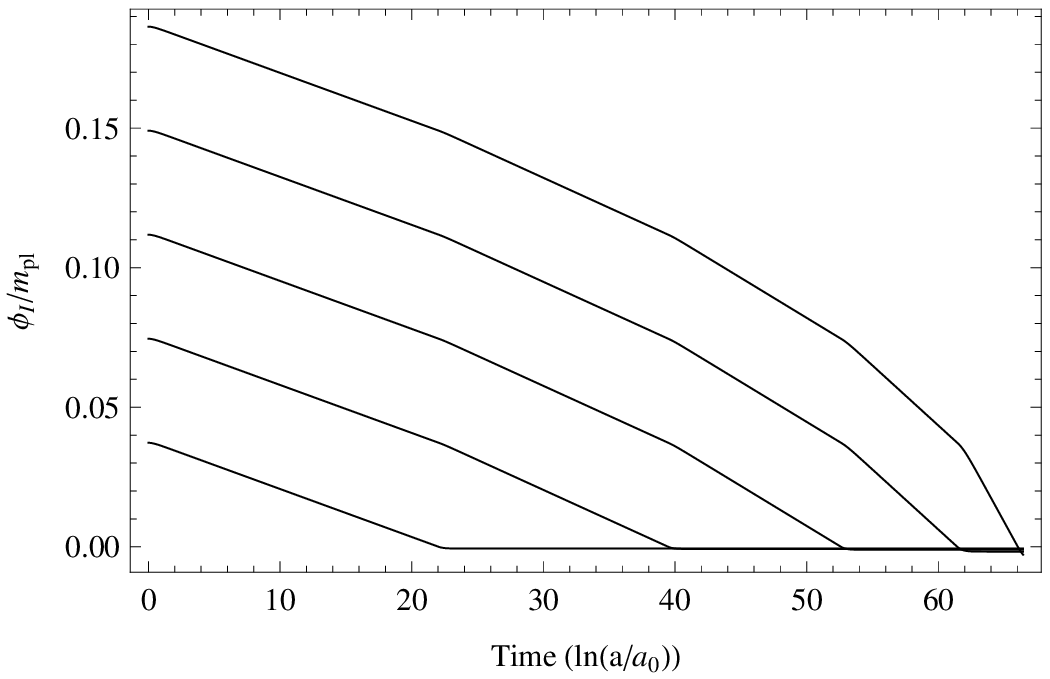}} &

\subfloat[]{
\label{fig:1b}
\includegraphics[width= 0.35\textwidth]{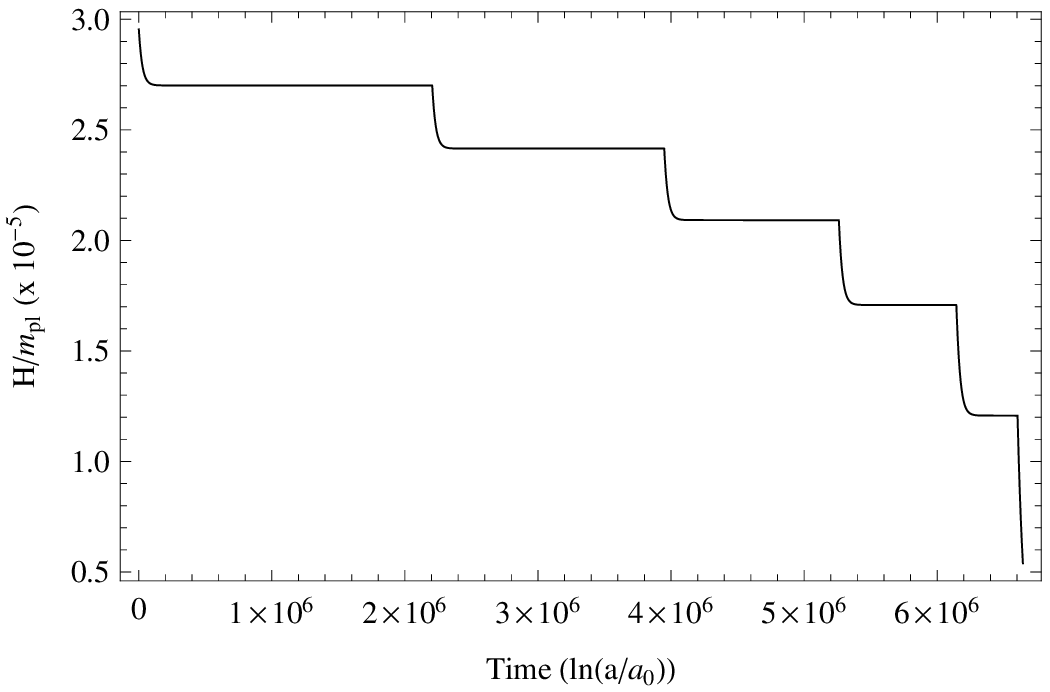}} \\

\subfloat[]{
\label{fig:1c}
\includegraphics[width= 0.35\textwidth]{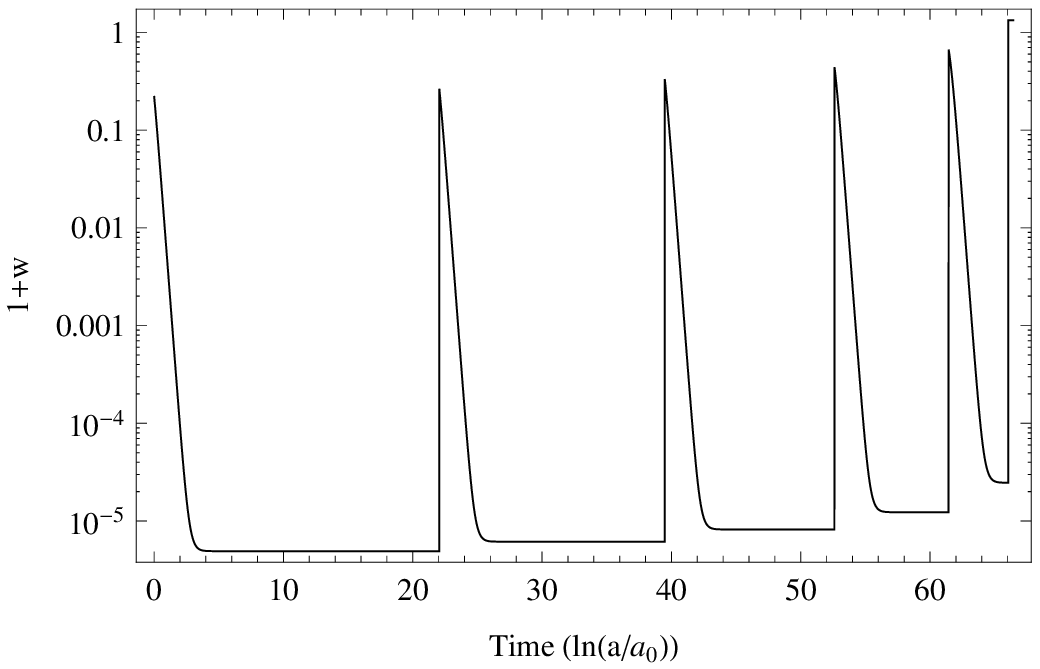}} &

\subfloat[]{
\label{fig:1d}
\includegraphics[width= 0.35\textwidth]{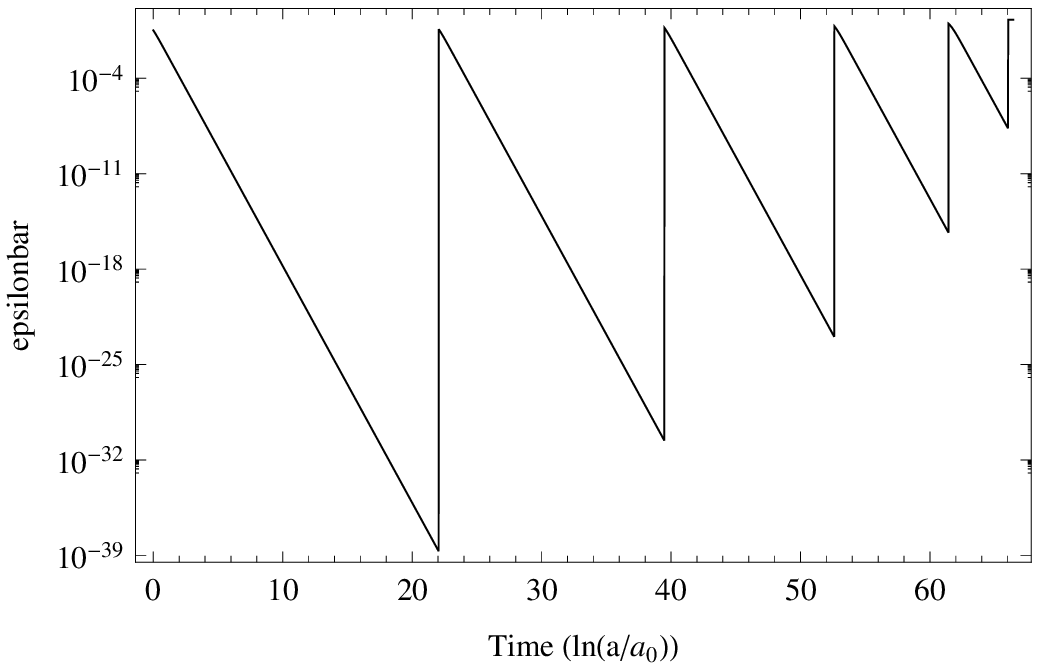}} \\

 \subfloat[]{
\label{fig:1e}
\includegraphics[width= 0.35\textwidth]{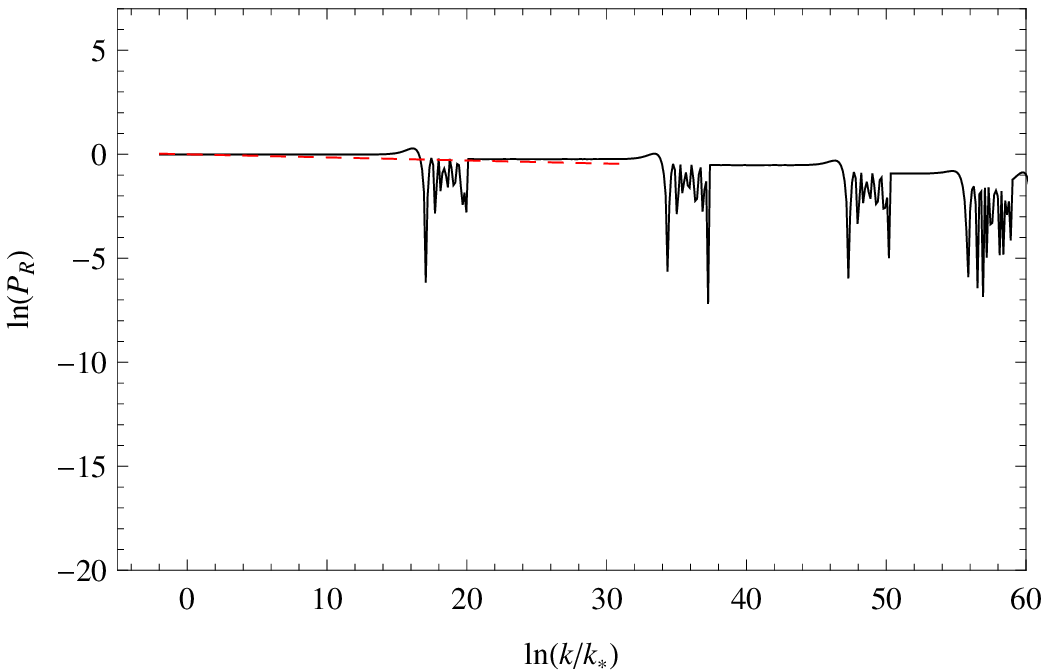}} &

\subfloat[]{
\label{fig:1f}
\includegraphics[width= 0.35\textwidth]{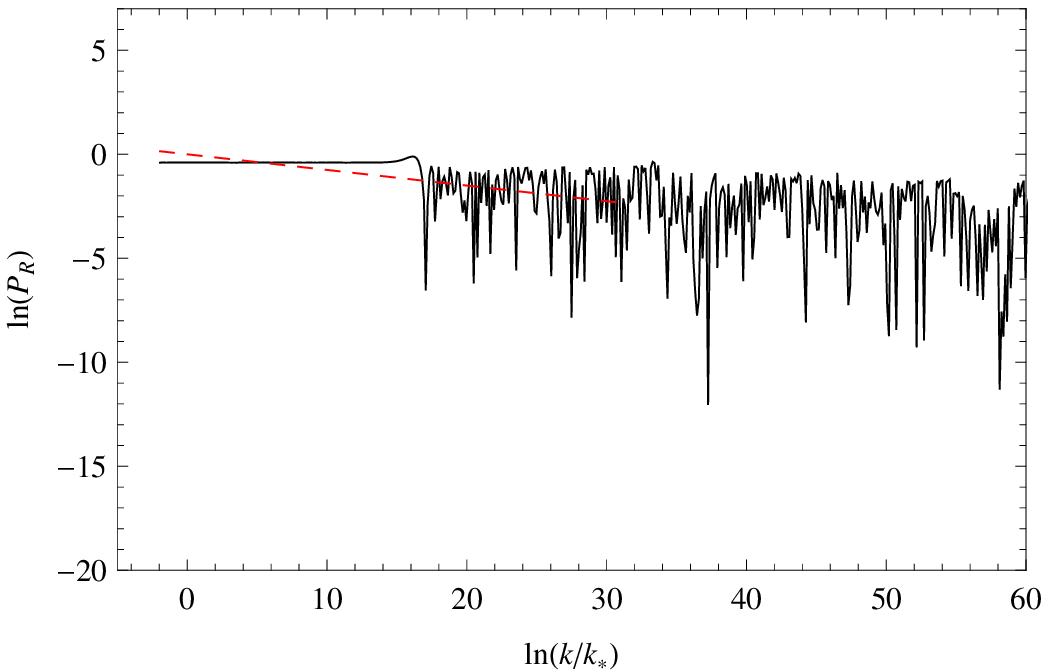}} \\

\end{tabular}

\caption{
The first four panels are (a) the evolution of each $\phi_I$, (b) the Hubble parameter, $H/m_{rm pl}$, (c) the equation of state $1+w$ and (d) $\bar{\epsilon}$ as a function of time throughout the simulation for $\mathcal{N}=5$.  The final two panels show the power spectrum when (e) the Bogoliubov coefficients are fixed and (f) Bogoliubov are matched every time a field drops out.  The simulation begins with $a_0=1$ and the power spectra are normalized to unity when $k_*$ is horizon sized 60 e-folds before the end of inflation.\label{fig1}
}
\end{figure}

We begin by simulating a small number of fields, $\mathcal{N}=5$.  Fig.~\ref{fig1} shows the main quantities of interest from this simulation.

In Fig.~\ref{fig:1e} and Fig.~\ref{fig:1f}, we compare the effect of the window function onto the power-spectrum.
In Fig.~\ref{fig:1e} the Bogoliubov coefficients are fixed and in Fig.~\ref{fig:1f} they are matched every time a field drops out according to (\ref{alphanew}) and (\ref{betanew}). Fixed Bogoliubov coefficients correspond to a sharp window function at $k_{\mbox{\tiny num}}$. Without this window function, all modes within the horizon are affected by a field decay (an unphysical effect), which explains the oscillations in Fig.~\ref{fig:1f}. If the number of fields is large, slow-roll is hardly disturbed whenever a field encounters $\phi_{end}$ and whether or not a window function is used becomes irrelevant. We fix the Bogoliubov coefficients for the rest of our simulations--utilizing a sharp window function.

\begin{figure}[tb]
 \begin{tabular}{cc}
\subfloat[]
{\label{fig:2a}
\includegraphics[width= 0.35\textwidth]{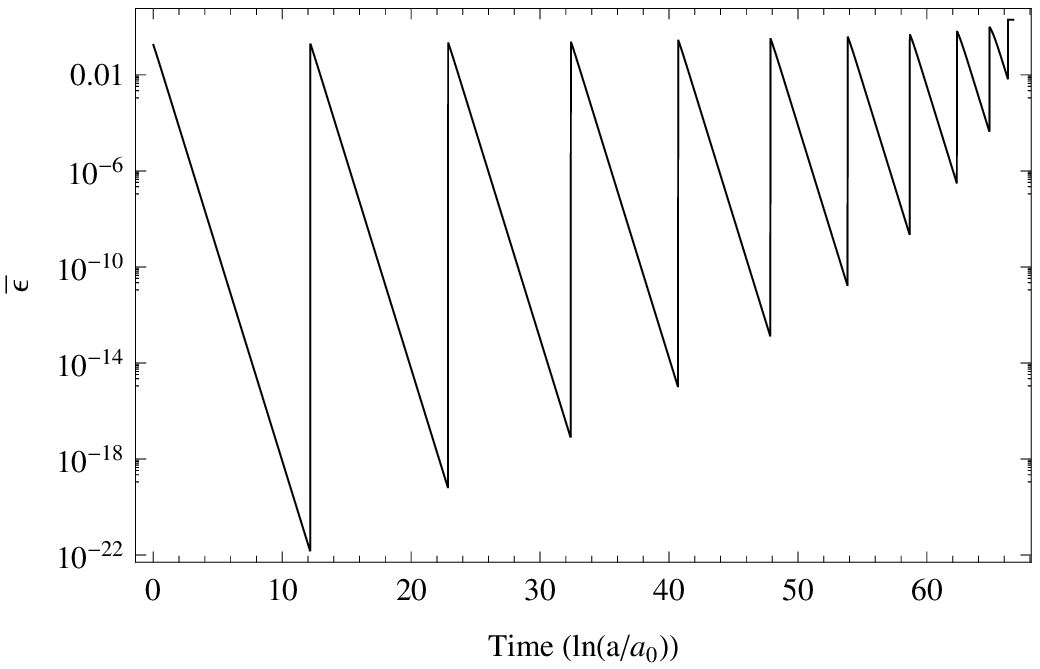}} &

\subfloat[]{
\label{fig:2b}
\includegraphics[width= 0.35\textwidth]{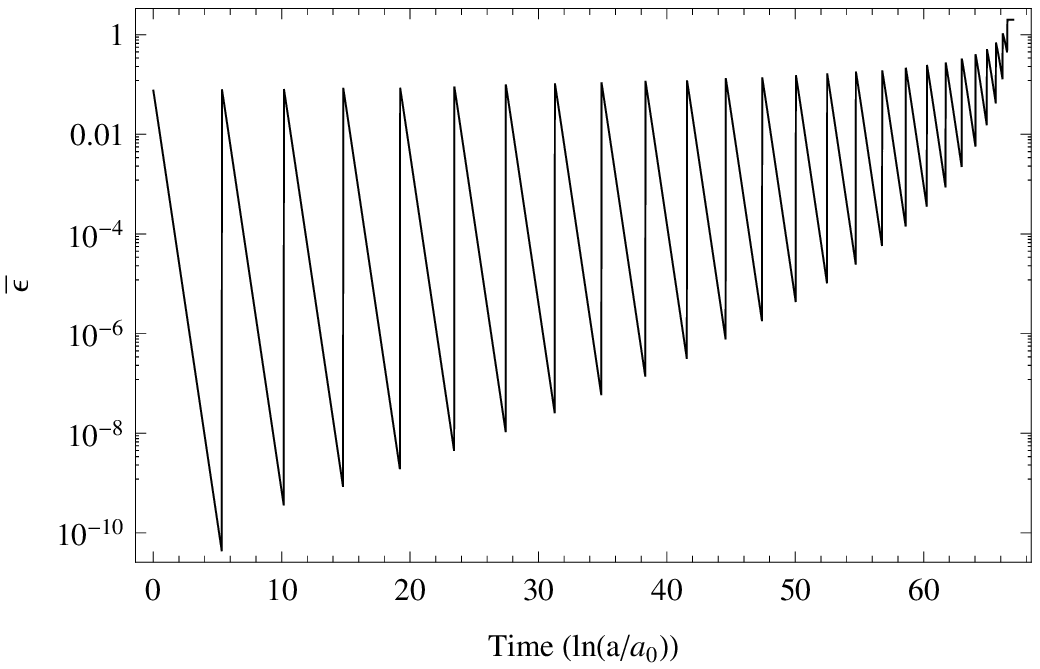}} \\

\subfloat[]{
\label{fig:2c}
\includegraphics[width= 0.35\textwidth]{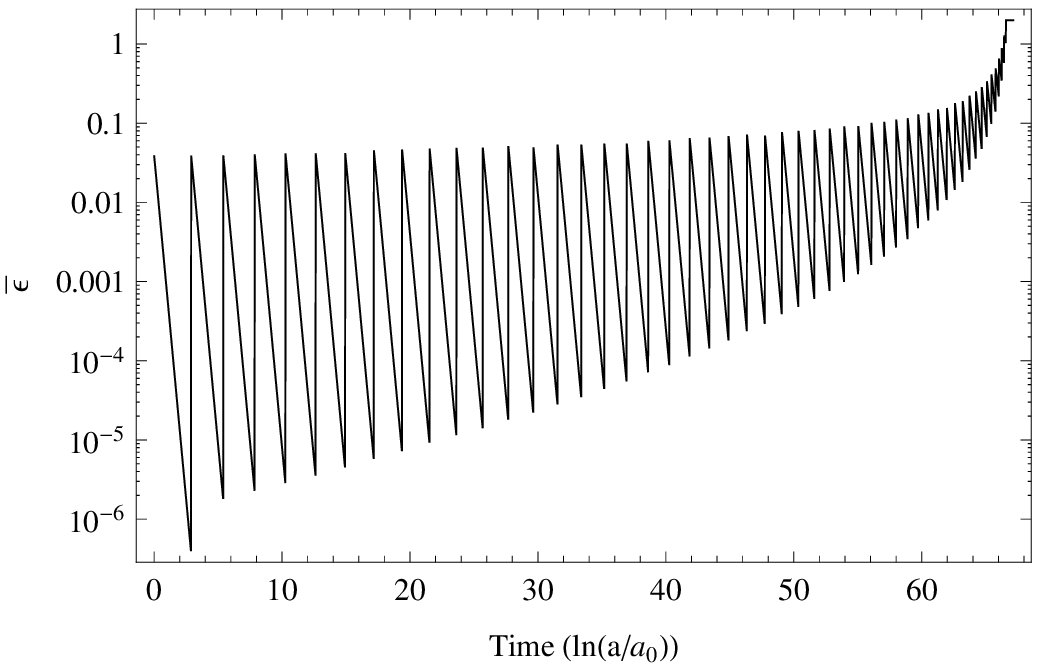}} &

\subfloat[]{
\label{fig:2d}
\includegraphics[width= 0.35\textwidth]{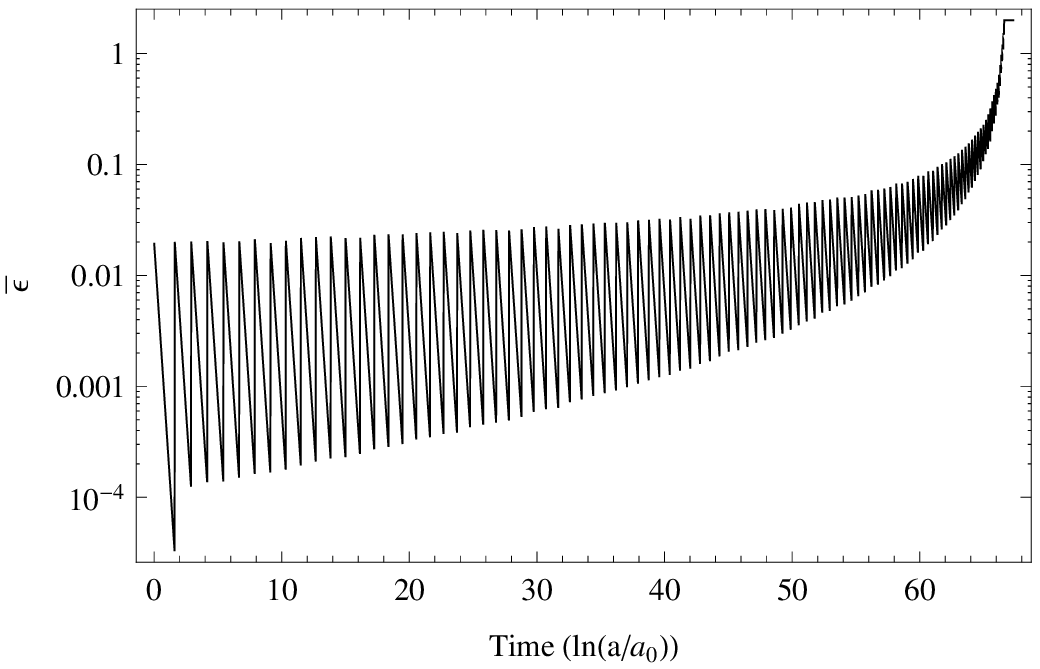}} \\

 \subfloat[]{
\label{fig:2e}
\includegraphics[width= 0.35\textwidth]{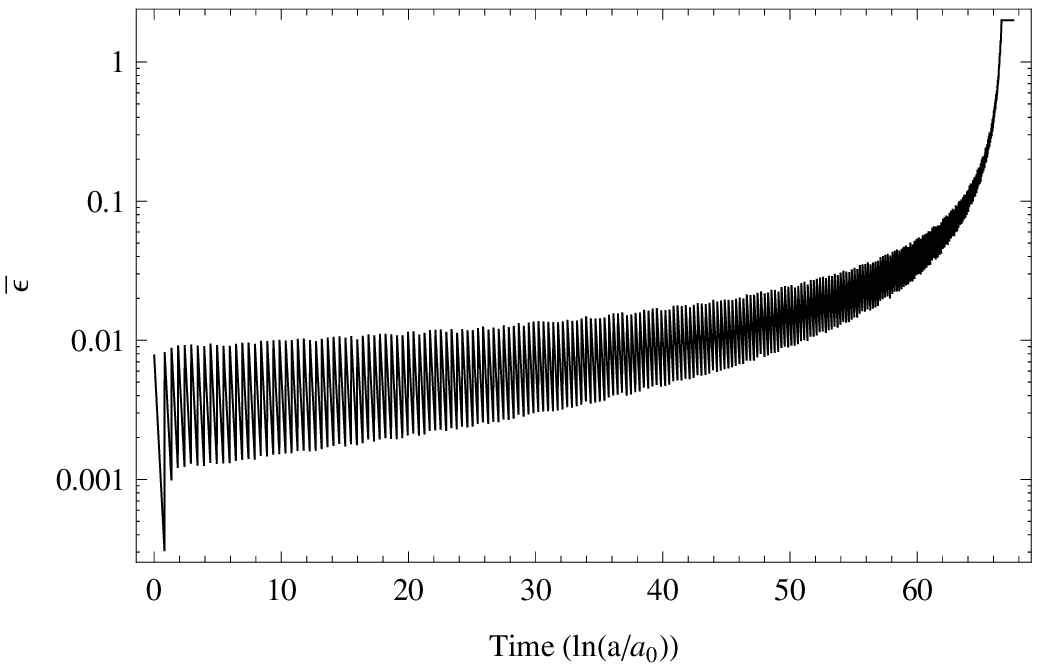}} &

\subfloat[]{
\label{fig:2f}
\includegraphics[width= 0.35\textwidth]{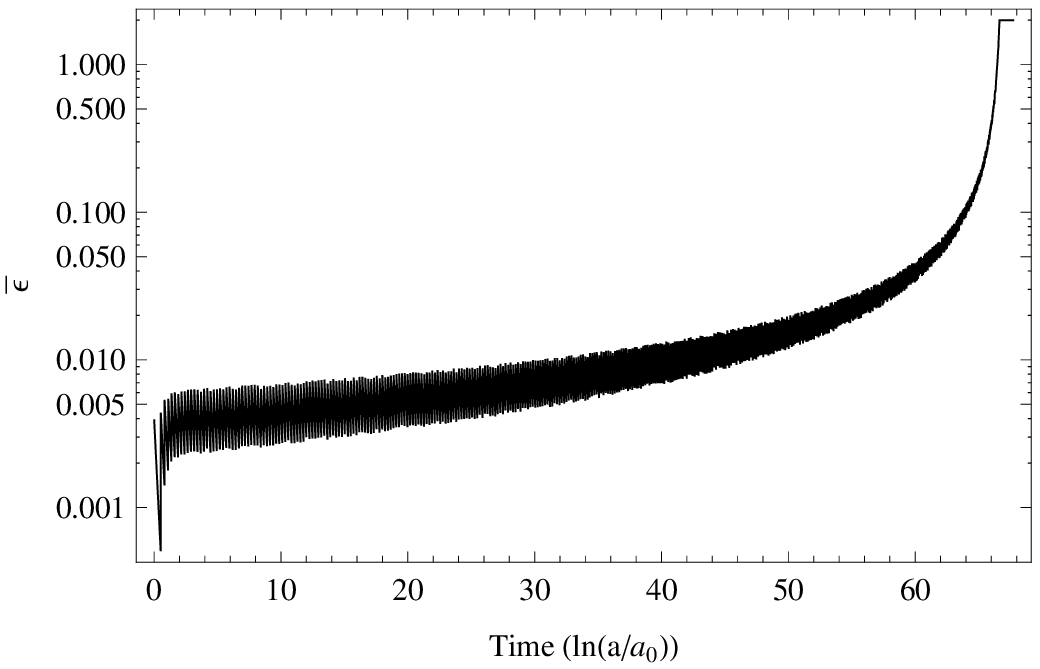}} \\

\end{tabular}

   \caption{The time evolution of $\bar{\varepsilon}=2\rho_r/(\rho_r+\rho_{\mbox{\tiny inf}})$ is plotted, using the number of e-folds as a time variable. The number of fields is varied, a) $\mathcal{N}=10$, (b) $\mathcal{N}=25$, (c) $\mathcal{N}=50$, (d) $\mathcal{N}=100$, (e)$\mathcal{N}=250$, (f) $\mathcal{N}=500$. For small $\mathcal{N}$, $\bar{\varepsilon}$ changes drastically. For large $\mathcal{N}$, radiation approaches a scaling regime and the validity range of the analytic approach to staggered inflation in \cite{Battefeld:2008py,Battefeld:2008qg} is entered.
\label{fig2}}
\end{figure}

\begin{figure}[tb]
 \begin{tabular}{cc}
\subfloat[]
{\label{fig:3a}
\includegraphics[width= 0.35\textwidth]{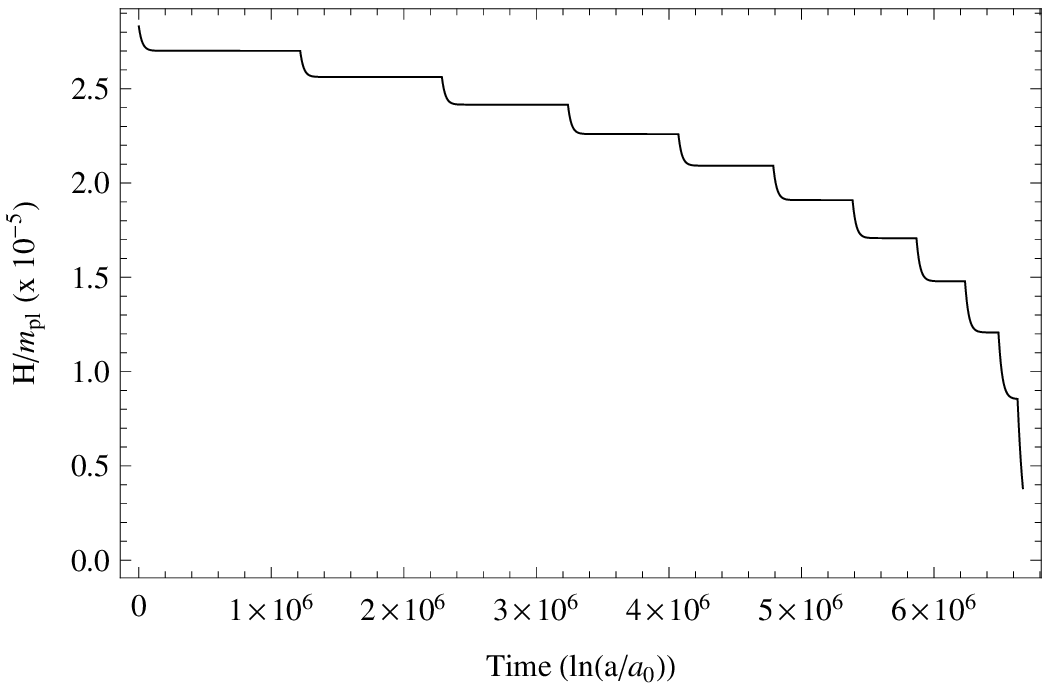}} &

\subfloat[]{
\label{fig:3b}
\includegraphics[width= 0.35\textwidth]{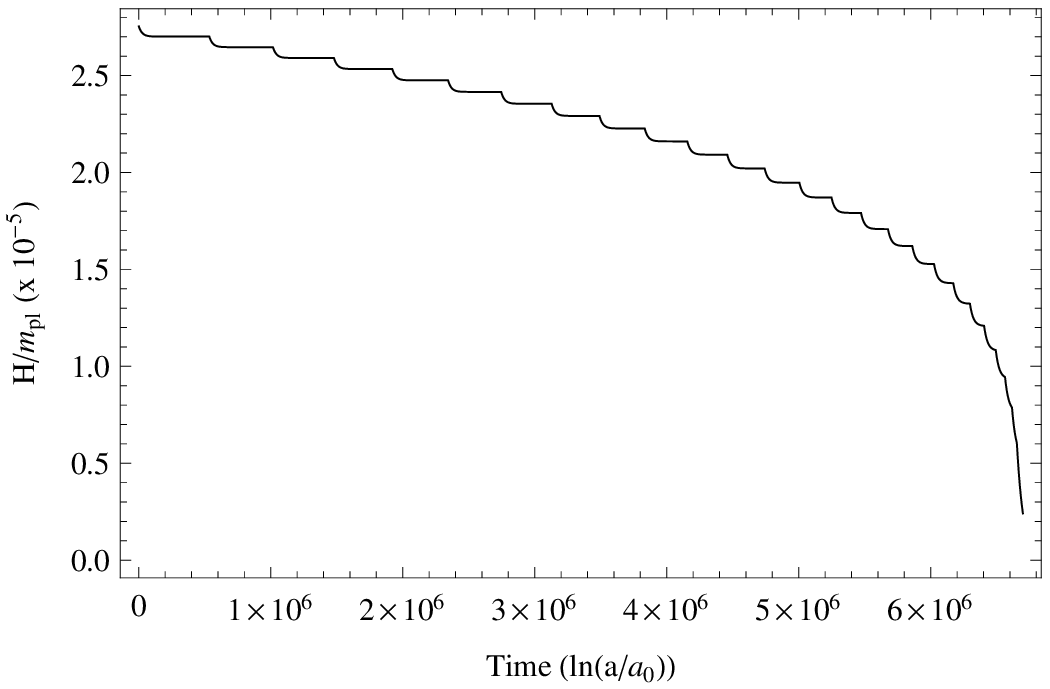}} \\

\subfloat[]{
\label{fig:3c}
\includegraphics[width= 0.35\textwidth]{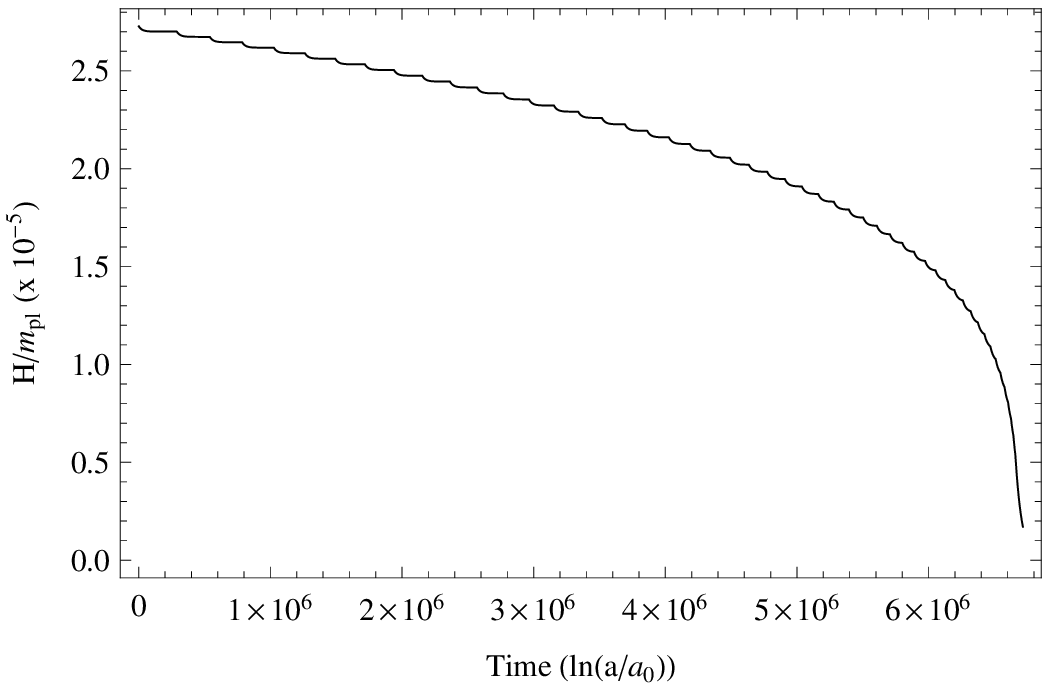}} &

\subfloat[]{
\label{fig:3d}
\includegraphics[width= 0.35\textwidth]{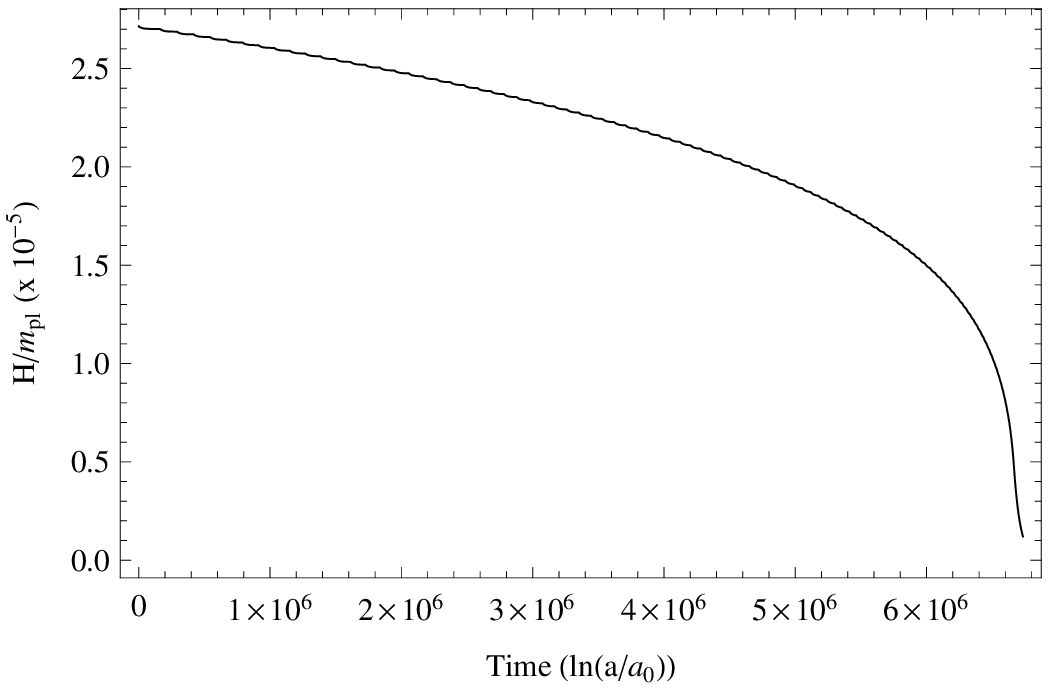}} \\

 \subfloat[]{
\label{fig:3e}
\includegraphics[width= 0.35\textwidth]{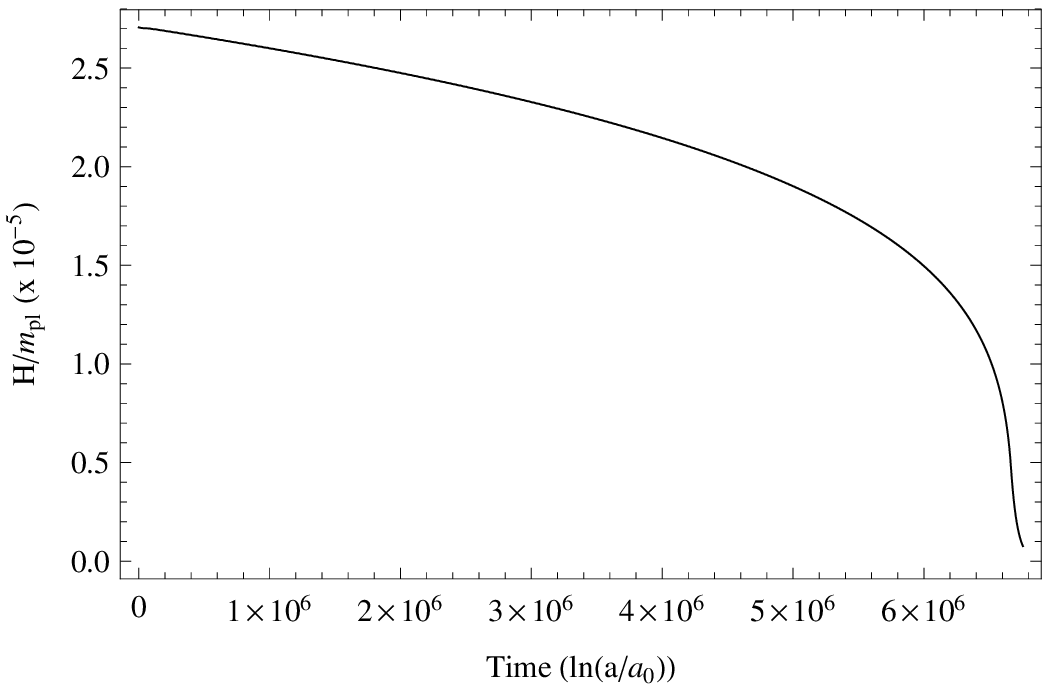}} &

\subfloat[]{
\label{fig:3f}
\includegraphics[width= 0.35\textwidth]{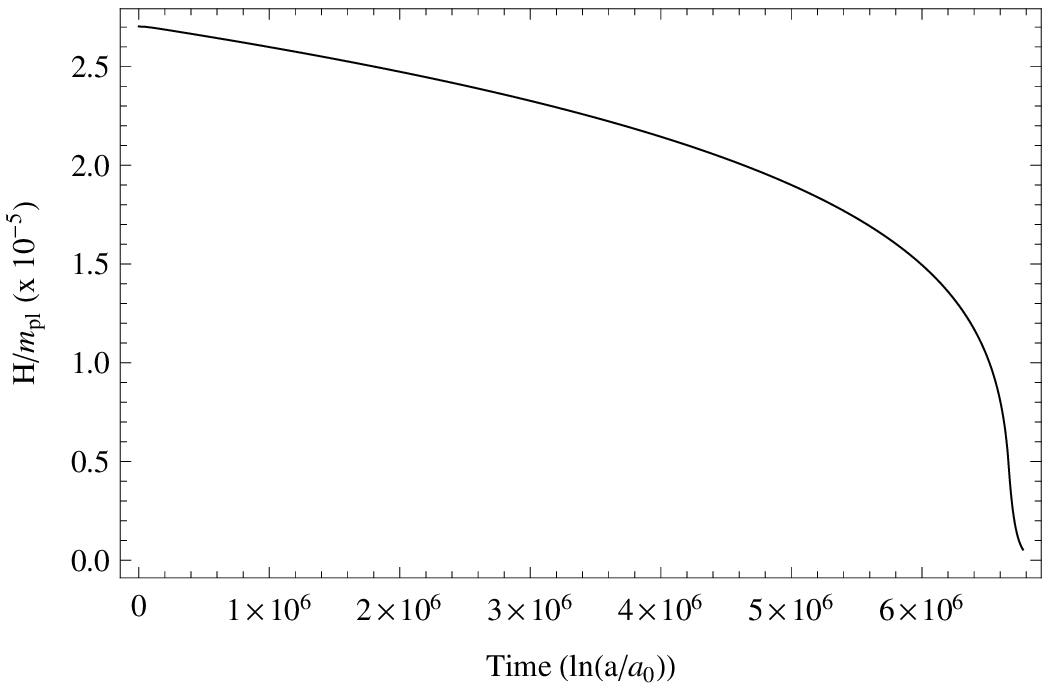}} \\

\end{tabular}

   \caption{The time evolution of the Hubble parameter is plotted, using the number of e-folds as a time variable. The number of fields is varied a) $\mathcal{N}=10$, (b) $\mathcal{N}=25$, (c) $\mathcal{N}=50$, (d) $\mathcal{N}=100$, (e)$\mathcal{N}=250$, (f) $\mathcal{N}=500$. For small $\mathcal{N}$, step-like features are evident, which wash out once $\mathcal{N}$ is increased. For large $\mathcal{N}$, the validity range of the analytic approach to staggered inflation in \cite{Battefeld:2008py,Battefeld:2008qg} is approached as the evolution of the Hubble parameter becomes smooth.
\label{fig3}}
\end{figure}

\begin{figure}[tb]
 \begin{tabular}{cc}
\subfloat[]
{\label{fig:4a}
\includegraphics[width= 0.35\textwidth]{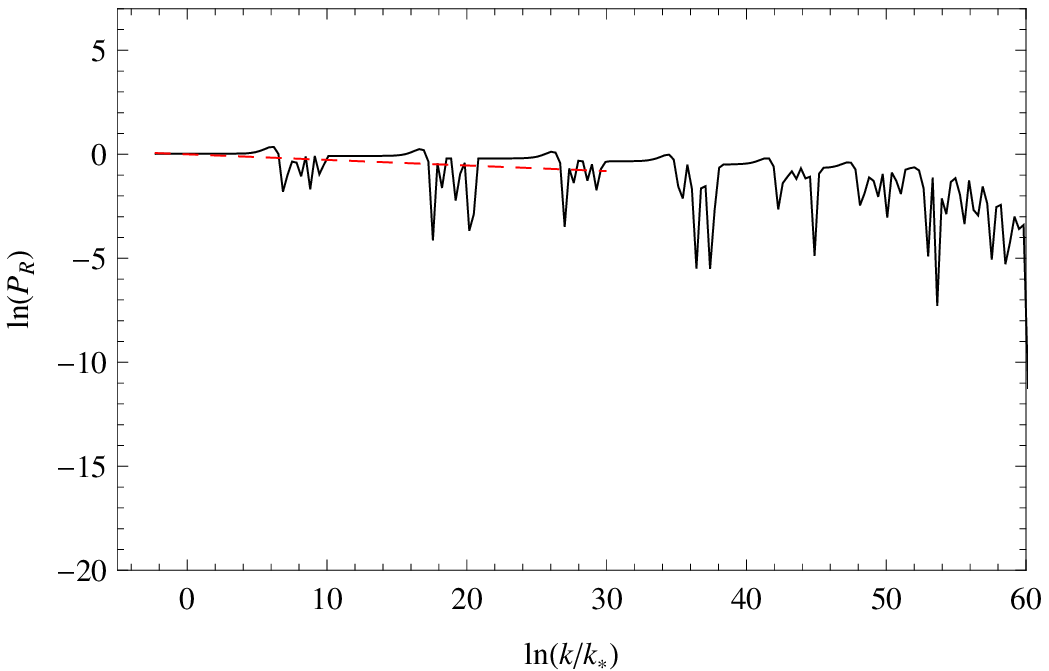}} &

\subfloat[]{
\label{fig:4b}
\includegraphics[width= 0.35\textwidth]{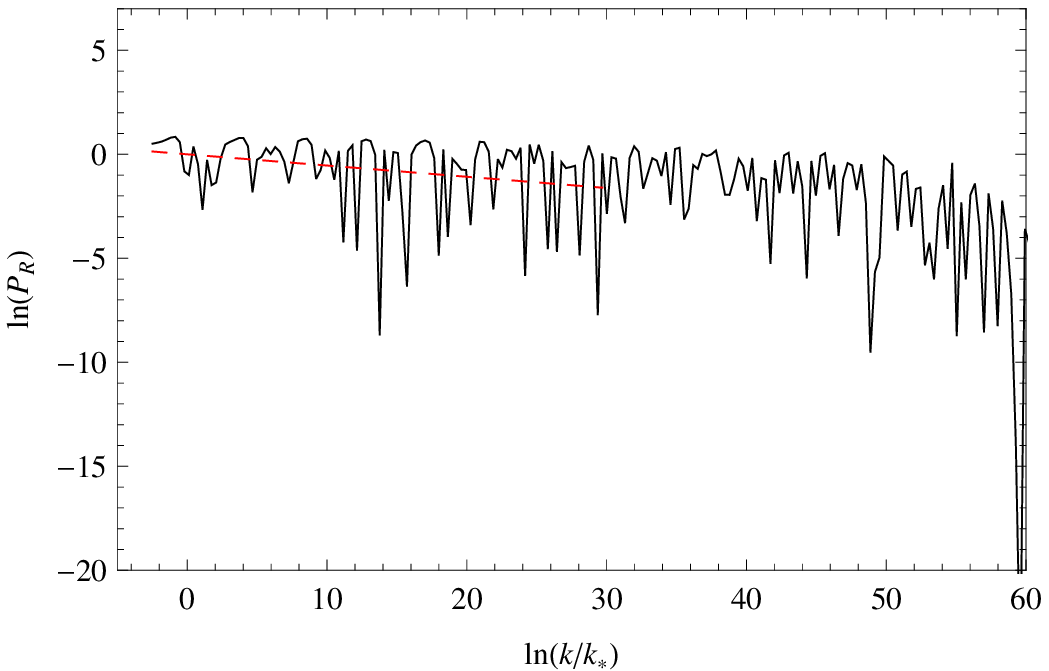}} \\

\subfloat[]{
\label{fig:4c}
\includegraphics[width= 0.35\textwidth]{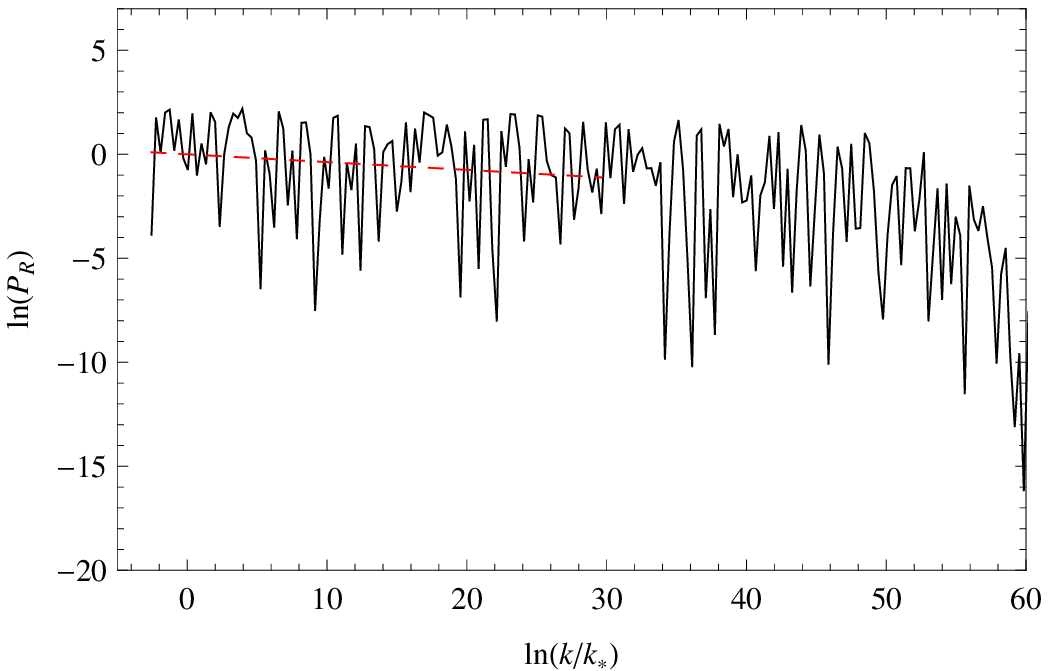}} &

\subfloat[]{
\label{fig:4d}
\includegraphics[width= 0.35\textwidth]{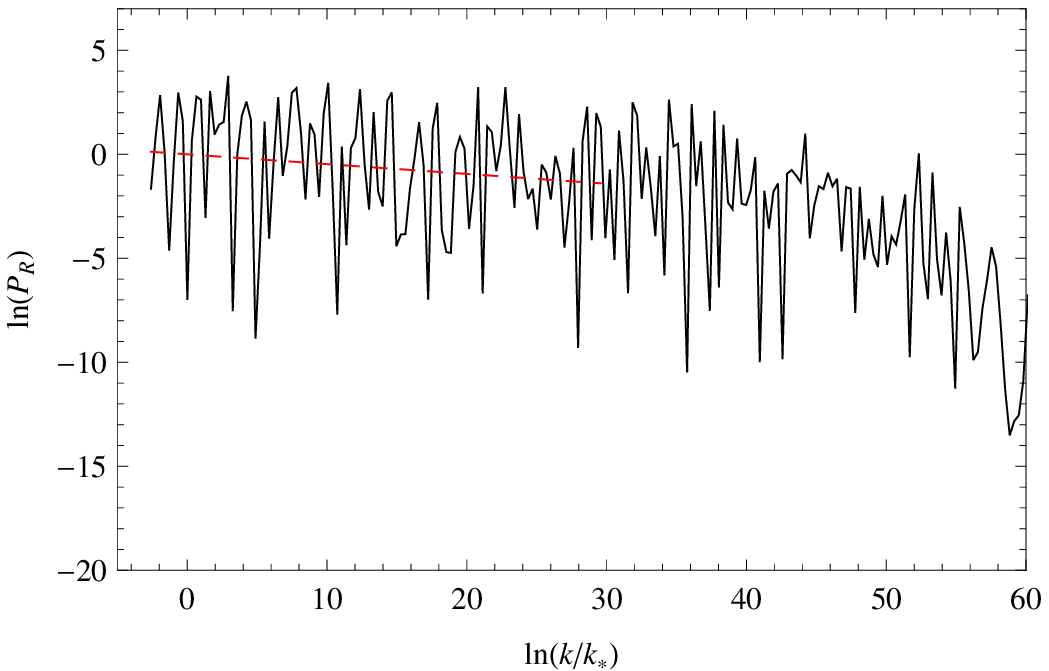}} \\

 \subfloat[]{
\label{fig:4e}
\includegraphics[width= 0.35\textwidth]{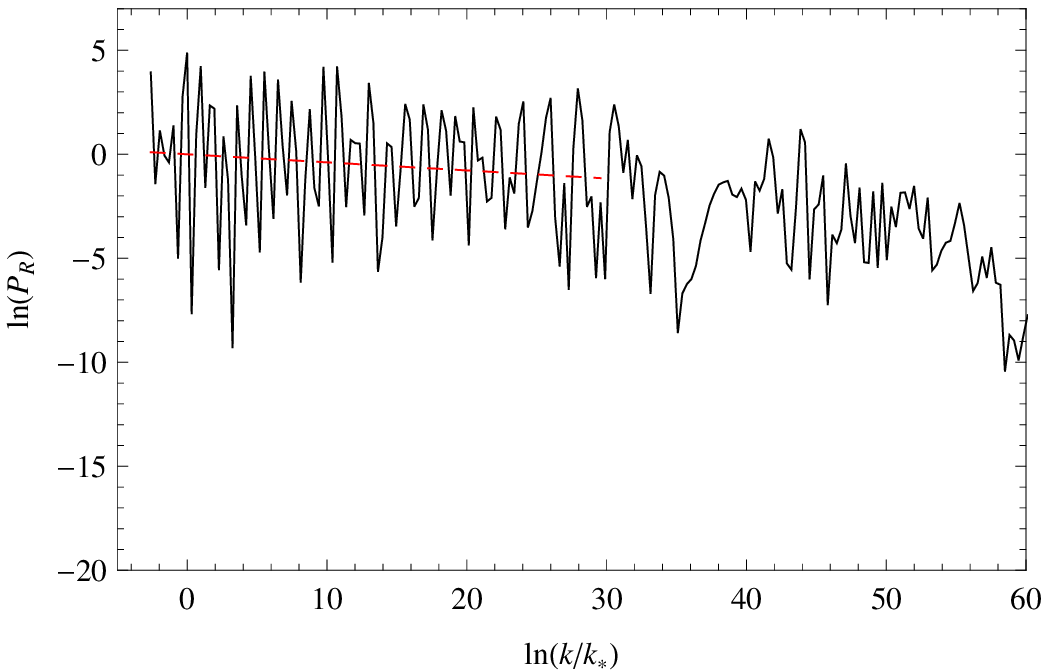}} &

\subfloat[]{
\label{fig:4f}
\includegraphics[width= 0.35\textwidth]{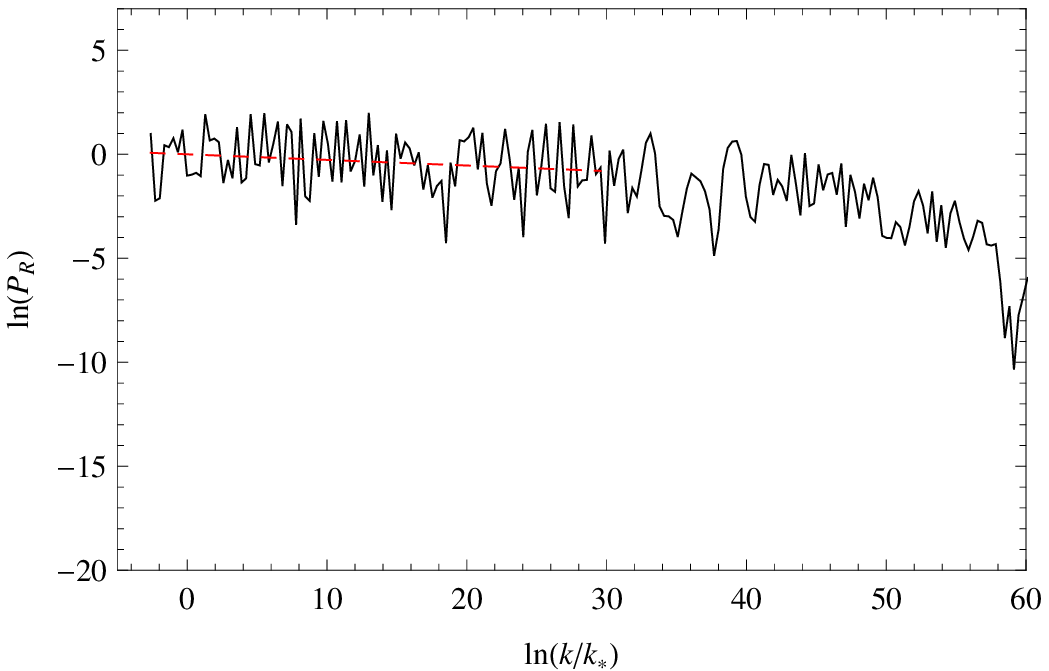}} \\

\end{tabular}

   \caption{The power-spectrum evaluated numerically towards the end of inflation is plotted over the wave-number for (a) $\mathcal{N}=10$, (b) $\mathcal{N}=25$, (c) $\mathcal{N}=50$, (d) $\mathcal{N}=100$, (e)$\mathcal{N}=250$, (f) $\mathcal{N}=500$. Signatures of individual decays wash out with an increasing number of fields.
\label{fig4}}
\end{figure}

\begin{figure}[tb]
 \begin{tabular}{cc}
\subfloat[]
{\label{fig:5a}
\includegraphics[width=0.35\textwidth]{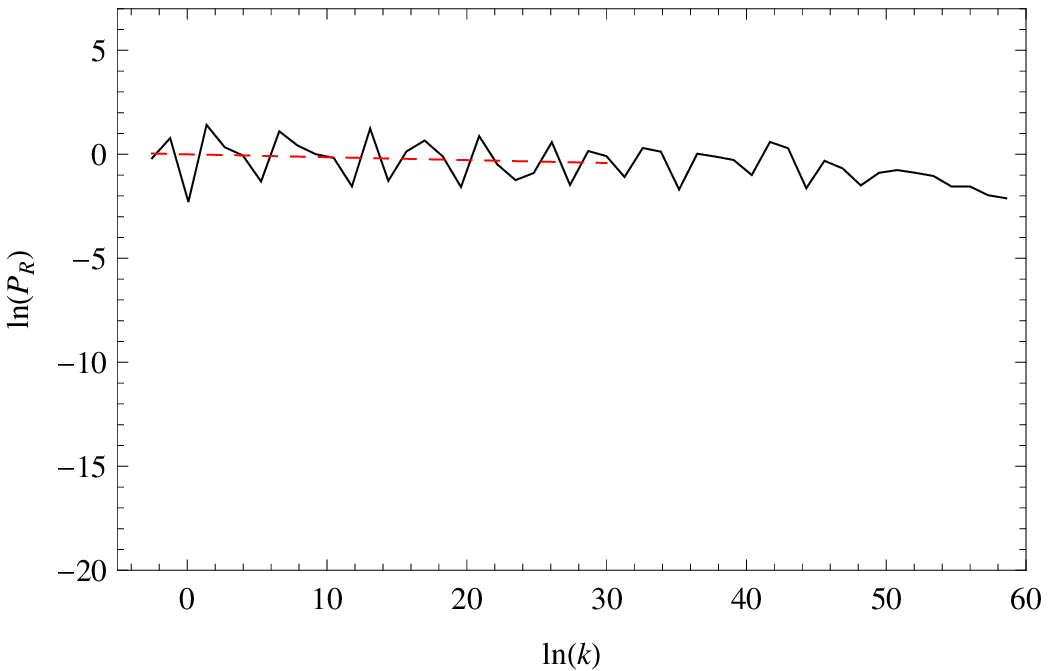}} &

\subfloat[]{
\label{fig:5b}
\includegraphics[width=0.35\textwidth]{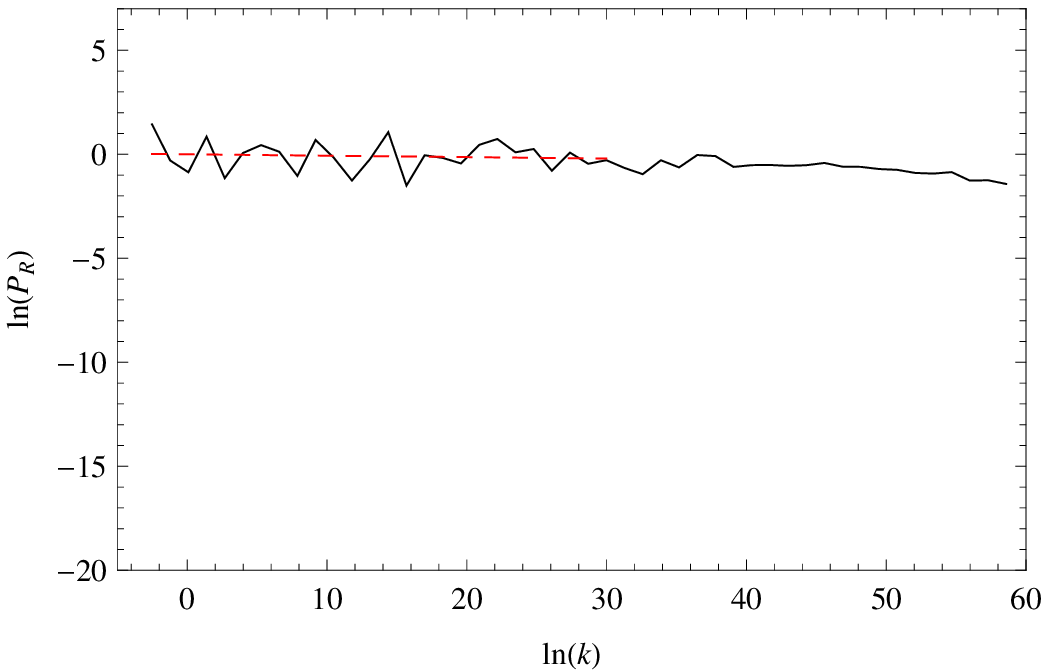}} \\

\subfloat[]{
\label{fig:5c}
\includegraphics[width= 0.35\textwidth]{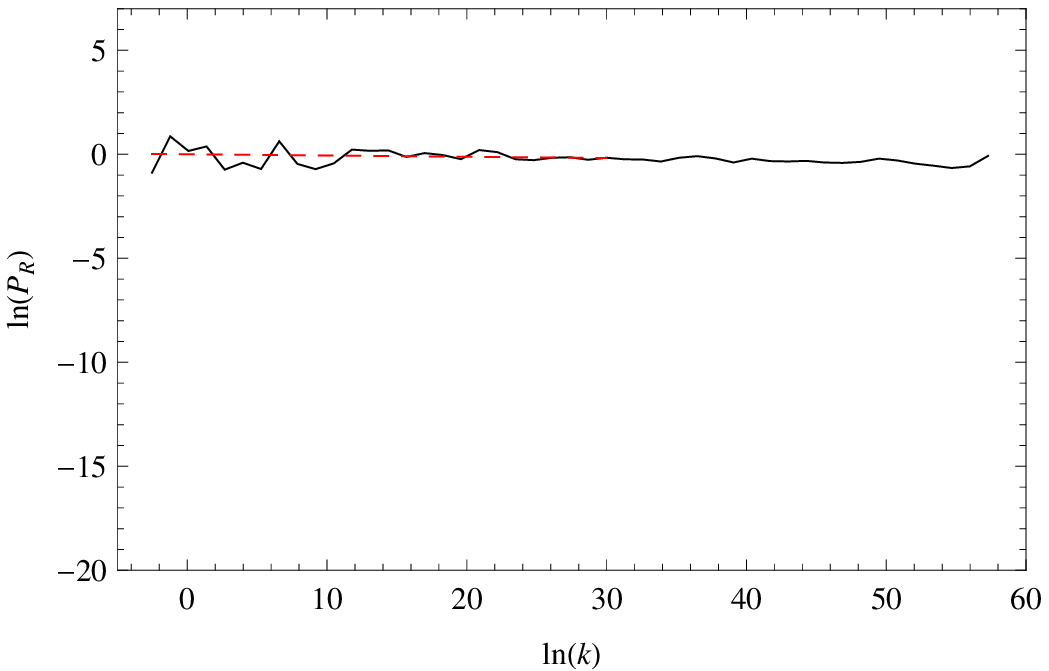}} &

\subfloat[]{
\label{fig:5d}
\includegraphics[width= 0.35\textwidth]{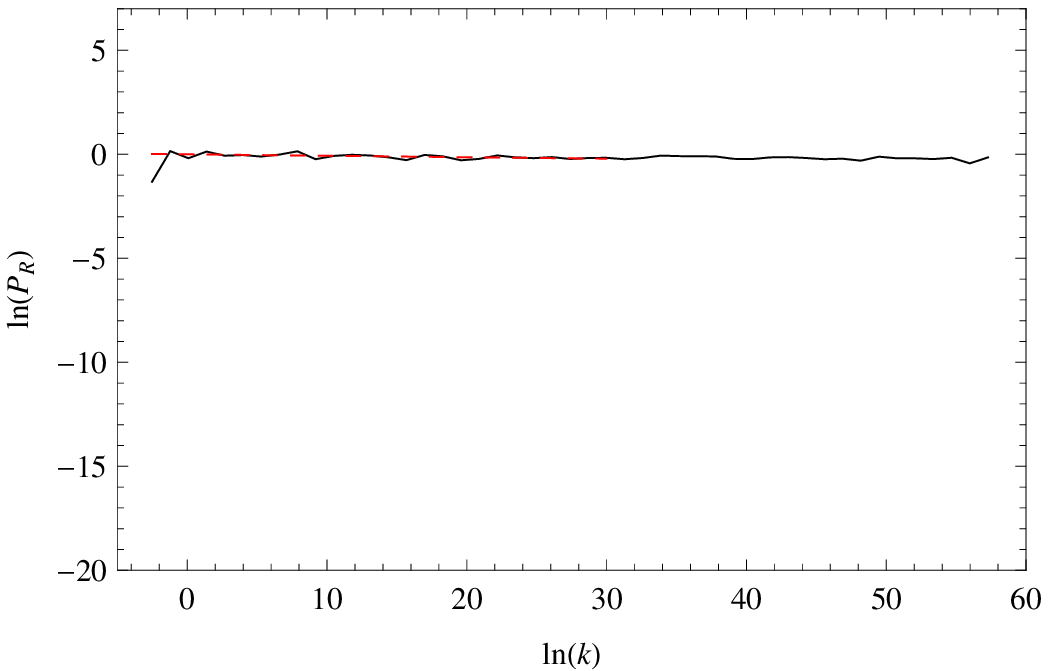}}\end{tabular}

   \caption{The power-spectrum evaluated numerically towards the end of inflation is plotted over the wave-number for (a) $\mathcal{N}=1000$, (b) $\mathcal{N}=1500$, (c) $\mathcal{N}=2000$, (d) $\mathcal{N}=2500$. These simulations involve considerably fewer modes than the models depicted in Fig.~\ref{fig5}.
\label{fig5}}
\end{figure}

In Fig.\ref{fig2} we plot $\bar{\epsilon}$ over the number of efolds for varying $\mathcal{N}$ and in Fig.~\ref{fig3} we plot the Hubble parameter, $H$, for the same set of simulations. For small $\mathcal{N}$, we observe a step-like reduction of $H$ whenever a field encounters a drop. These steps are smoothed out, since we track the radiation into which the fields' potential energy is infused. This goes hand in hand with large changes in $\bar{\epsilon}$, Fig.~\ref{fig2}. The presence of these steps, and the accompanying change in the equation of state parameter, give rise to ringing patterns in the power-spectrum for large $k$. For low $\mathcal{N}$, these ringing patterns are problematic due to their large amplitude. Away from the decay events, the power-spectrum becomes smooth again, as expected during slow roll. As the number of fields is increased, the steps begin to overlap, leading to a smooth reduction of the inflationary energy and less variations in $\bar{\epsilon}$; as a consequence, features in the power-spectrum overlap, leading to a smooth spectrum in the large $\mathcal{N}$ limit, see Fig.~\ref{fig4} and Fig.~\ref{fig5}. 

A word of caution is in order here: in our plots, we rescaled the amplitude of the power-spectrum to one at sixty efolds before the end of inflation. The actual amplitude increased in our numerical code (independent of k) as the number of fields increases. This effect is due to the rescaling of the Bogoliubov coefficients whenever a field drops out and thus an artifact of ignoring perturbations in radiation. The shape of the spectrum is entirely unaffected.

The presence of ringing patterns at large $k$ as found in \cite{Battefeld:2010rf} is easy to understand, since the power-spectrum is modulated by the combination $|\alpha-\beta|^2$ of the Bogoliubov coefficients. If we consider the case of a single decay ($\beta_I^-=0$ and $\alpha_I^-=1$), work in zeroth order of small parameters  ($A_I\rightarrow 0$), and ignore the window function in (\ref{windowfunction}) we arrive at a modulation of the I'th field's power-spectrum of  
\begin{eqnarray}
\left| \alpha_I^+ -\beta^I_+ \right|^2 &\simeq& \frac{1}{B}+\frac{B-1}{B}\left(1-\cos (2x)-\frac{1}{x}\sin (2x)\right)\\
&&+\frac{(B-1)^2}{2B}\left(1-\cos (2x)\left(1+\frac{1}{x^2}\right)\right)\,,
\end{eqnarray}
where we used (\ref{alpha}) and (\ref{beta}), rescaled the Bogoliubov coefficients with $1/\sqrt{B}$ to guarantee $|\alpha_I^2|-|\beta_I|^2=1$, and defined $x\equiv k/\mathcal{H}_{\mbox{\tiny decay}}$ in terms of the comoving mavenumber $k$ and the Hubble scale at the time of the transition $\mathcal{H}_{\mbox{\tiny decay}}$. Only modes within the horizon at the time of the transition pick up these modulations; if more fields subsequently decay, the patterns get more complicated and ultimately wash out. Furthermore, the oscillations are damped away once $k$ approaches the inverse of the decay time. 

Another word of caution: since we used $\left[\mathcal{R}\right]_\pm =0$ in deriving the above, we ignored a term proportional to $x^2=(k\mathcal{H})^2$; if this term were kept, additional contributions would result for large $x$ that scale as $x^2$, dominating over the corrections we kept in the large $k$ limit. Thus we provide a lower bound on features in the power-spectrum by using $\left[\mathcal{R}\right]_\pm =0$ and the resulting matching conditions.

For $\mathcal{N}\sim 10^3$, we start to recover the analytic results of \cite{Battefeld:2008py,Battefeld:2008qg}, as evident by Fig.~\ref{fig4} and \ref{fig5} and a comparison of the scalar spectral index found numerically to (\ref{nslinear}), see table \ref{table1}. To compute the spectral index in table \ref{table1}, we average the slope of the 
power-spectrum in Fig.~\ref{fig4} and \ref{fig5} between $60$ and $30$ e-folds before the end of inflation. Naturally, such a smoothing is not a good approximation for the spectrum if pronounced patterns are present, as in Fig.~\ref{fig1} and Fig.~\ref{fig2} for low $\mathcal{N}$. Thus the low $\mathcal{N}$ values for $n_s$ should be taken with caution. Since the running time of our code scales linearly with $\mathcal{N}$, and we track several hundred modes per field\footnote{200 modes are tracked in the simulations in Fig.~\ref{fig4} and 50 modes are tracked in the simulations in Fig.~\ref{fig5}.}, we were currently limited to $\sim 10^3$ fields (a run takes around a week).  

\begin{table}[htdp]
\begin{center}
\begin{tabular}{|c|c|}
\hline
$\mathcal{N}$ & $n_s-1$\\ 
\hline
\hline
$5$ & $-0.0149$\\ 
\hline
$10$ & $-0.0269$\\ 
\hline
$25$ & $-0.0540$\\ 
\hline
$50$ & $-0.0375$\\ 
\hline
$100$ & $-0.0472$\\ 
\hline
$250$ & $-0.0388$\\ 
\hline
$500$ & $-0.0270$ \\ 
\hline
$1000$ & $-0.0165$ \\ 
\hline
$1500$ &  $-0.00676$ \\ 
\hline
$2000$ & $-0.00598$ \\ 
\hline
$2500$ & $-0.00707$ \\ 
\hline\hline
$\infty$ & $-0.0167$\\ 
\hline
\end{tabular} 
\caption{Comparison of the scalar spectral indices found numerically for varying number of fields to the asymptotic $\mathcal{N}=\infty$ result in (\ref{nslinear}), $n_s-1\simeq-1/N=-1/60\approx -0.0167$, based on the analytic treatment of staggered inflation in Sec.~\ref{sec:largeNlimit}, \cite{Battefeld:2008py,Battefeld:2008qg}. For low $\mathcal{N}$ features in the power-spectrum are so pronounced that a power-law is not a good approximation and the values for $n_s$ should be taken with caution. The difference in the large $\mathcal{N}$ limit is caused by our neglection of perturbations in $\rho_r$ in our code, which are accounted for in the analytic approach.  \label{table1}}
\end{center}
\end{table}

We see that the analytic approach to staggered inflation in \cite{Battefeld:2008py,Battefeld:2008qg} becomes a good approximation if  around $40$ or more fields drop out in a given Hubble time. In this case the evolution of $H$ is sufficiently smooth when modes relevant to observations leave the horizon and features due to individual decays in the power-spectrum are washed out.

To summarize, from our comparisons we can draw two conclusions; firstly, the analytic results of \cite{Battefeld:2008py,Battefeld:2008qg} can be recovered numerically in the large $\mathcal{N}$-limit. Furthermore, the distinct features for low $\mathcal{N}$ are in line with \cite{Battefeld:2010rf}. Secondly, our numerical code reliably tracks the evolution of perturbations in models with decaying fields. 

The main simplifying approximations we made were the neglection of perturbations in radiation and the use of a sudden decay approximation, requiring the inclusion of a window function (\ref{windowfunction}) in the matching conditions to induce a suppression of the ringing patterns on small scales.  Any surviving ringing patterns on small scales should be taken with caution, since we ignored terms proportional to $(k/\mathcal{H})^2$ in deriving the matching conditions for perturbations.\footnote{Note that in \cite{Firouzjahi:2010ga} the terms of order $\mathcal{O}((k/\mathcal{H})^0)$, which we keep, are ignored.} 

We focused on a simple multi-field model, that, although motivated by moduli dynamics on the landscape, should primarily be seen as a toy model to test the tools developed so far to deal with staggered inflation.

\section{Conclusion}
We computed numerically the scalar power-spectrum in a simple multi-field inflationary setup that contains decaying fields during inflation, varying the number of fields from a few to thousands, motivated by multi-field models in string theory. For low field numbers, we find distinguishable patterns in the power-spectrum caused by individual decays, in line with the analytic results of \cite{Battefeld:2010rf}, where a single decay has been investigated in a concrete string-derived model. Once many fields decay during any given Hubble time, we recover the results of the analytic treatment of staggered inflation in \cite{Battefeld:2008py,Battefeld:2008qg}: the superposition of many patterns yields a smooth spectrum that is well described by a power-law; the scalar spectral index is not only set by the slow roll parameters, but also by the ratio of the decay rate to the Hubble parameter which, in the model under consideration, is the dominant contribution.

The numerical confirmation of the analytic results in the large/small $\mathcal{N}$ limit, as well as the development of a reliable numerical code to deal with any number of fields are the main results of this paper.

A shortcoming of our current approach is the neglection of perturbations in the radiation bath that is produced whenever a field decays. This made an overall renormalization of the Bogoliubov coefficients a necessity whenever a field decayed and leads to order one deviations for the scalar spectral index as compared to the analytic results of staggered inflation. We do not expect qualitative differences if these perturbations were kept. We also ignored terms of order $k^2/\mathcal{H}^2$ and worked in a sudden decay approximation, which prevents a suppression of ringing patterns on scales deep within the horizon. To ameliorate this unphysical effect, we introduced a window-function into the matching conditions by hand, rendering modes deep inside the horizon oblivious to the fields' decay. 

We plan to use the code developed for this paper in an investigation of the extended KKLMMT setup in \cite{Battefeld:2010rf}, where we plan to relax these approximations.

\begin{acknowledgments}
We thank H.~Firouzjahi for discussions. Research at the Perimeter Institute for Theoretical Physics is supported by the Government of Canada through Industry Canada and by the Province of Ontario through the Ministry of Research \& Innovation.

\end{acknowledgments}

\appendix
\section{Code Implementation \label{sec:code}}
 
To evolve the scale factor computationally the Friedmann equation and Klein-Gordon field equation were rewritten as coupled first-order differential equations. These modified equations take the form
\begin{eqnarray}
\ddot{\phi_I} &=& -\frac{\partial V_I}{\partial\phi_I} - 3\dot{\xi}\theta_I \,, \\
\label{scalefactorEOM}
\dot{\xi} &=& \sqrt{\frac{8\pi G}{3}\left(\sum_I^{\mathcal{N}}\left( \frac{1}{2}\dot{\phi}^2_I+V_I \right) + \rho_r\right)}\,.
\end{eqnarray}
We track the natural logarithm of the scale factor, $\xi= \ln{a}$, which is the number of e-folds since the beginning of the simulation. 

Similarly, the second-order differential equation for the Sasaki-Mukhanov variable (\ref{analyticolQ}) is coupled to
(\ref{scalefactorEOM}) and can be decomposed into two coupled first-order differential equations. The coupled first-order Sasaki-Mukhanov equations are
\begin{eqnarray}
\dot{Q}_I &=& -3\dot{\xi}P_I- k^2e^{-2\xi}Q_I+8\pi G\sum^N_J\left[\left(\frac{\dot{\varphi}_I\ddot{\varphi}_J+\ddot{\varphi}_I\dot{\varphi}_J}{\dot{\xi}}+\left(4-\frac{\ddot{\xi}}{\dot{\xi}^2}\right)\dot{\varphi}_I\dot{\varphi}_J\right)Q_J\right] \,.
\label{evolveactual}
\end{eqnarray}
Since the Sasaki-Mukhanov variables are complex, we track both the real and imaginary components.  At the beginning of the simulation we chose a set of $k_1\ldots k_i$ where $i\approx 100-500$.   This set of discrete $Q_I(k_i)$ are initialized in the Bunch-Davies vacuum, (\ref{vacinit}) and the $Q_I(k_i)$ are identical for each $I$.  All modes are tracked analytically until the physical wavelength $a/k$ is some fraction $\beta$ of the Hubble length, i.e. until $k=aH/\beta$; for the simulations in this paper we take $\beta = \exp(-4)$.   At this point we evolve the $Q_I$ numerically according to (\ref{evolveactual}). 

The equations are evolved using the fifth-order Cash-Karp Runge-Kutta method of integration \cite{Press:1992}. The primary advantage of this method is its adaptive step size. At each time step the program computes values via fifth-order integration and an embedded fourth-order integration. The fastest oscillating mode in play determines the adjustable time-step.  All other modes in play oscillate at a lower frequency and are therefore not as sensitive to the timestep.

\end{document}